\documentclass[aps,prb,twocolumn,showpacs]{revtex4}

\usepackage{graphicx}
\usepackage{epsf} 
\usepackage{dcolumn}
\usepackage{bm}
\usepackage{amssymb}
\usepackage{amsmath}
\usepackage{amsbsy}
\usepackage{graphicx,color,ifthen}
\usepackage{pstricks,pst-node,pst-text,pst-3d,color}
\usepackage[latin1]{inputenc}

\newcommand{\Ge}{\Gamma_{ex}}
\newcommand{\Ga}{\Gamma_{ad}}

\newcommand {\Rwm} {{\rm Re} (\omega_{\mathbf{k}})}

\begin{document}


\title{Coupling of morphology to surface transport in ion-beam irradiated surfaces.
I. Oblique incidence}

\author{Javier Mu\~noz-Garc\'ia}
\email{javiermunozgarcia@gmail.com}

\homepage[]{http://gisc.uc3m.es/~javier}

\altaffiliation[Present address: ]{School of Mathematical Sciences and Complex and Adaptive Systems Laboratory,
University College Dublin, Belfield, Dublin 4, Ireland}

\affiliation{Grupo Interdisciplinar de Sistemas Complejos (GISC) and
Departamento de Matem\'aticas, Facultad de Ciencias Qu\'{\i}micas, Universidad
de Castilla-La Mancha, E-13071 Ciudad Real, Spain}

\author{Rodolfo Cuerno}
\affiliation{Departamento de Matem\'aticas and GISC, Universidad Carlos III de
Madrid, Avenida de la Universidad 30, E-28911 Legan\'es, Spain}

\author{Mario Castro}
\affiliation{GISC and Grupo de Din\'amica No Lineal (DNL), Escuela T\'ecnica
Superior de Ingenier{\'\i}a (ICAI), Universidad Pontificia Comillas de Madrid,
E-28015 Madrid, Spain}


\date{\today}

\begin{abstract}
We propose and study a continuum model for the dynamics of amorphizable surfaces undergoing ion-beam sputtering (IBS) at intermediate energies and oblique incidence. After considering the current limitations of more standard descriptions in which a single evolution equation is posed for the surface height, we overcome (some of) them by explicitly formulating the dynamics of the species that transport along the surface, and by coupling it to that of the
surface height proper. In this we follow recent proposals inspired by
``hydrodynamic'' descriptions of pattern formation in aeolian sand dunes and
ion-sputtered systems. From this enlarged model, and by exploiting the
time-scale separation among various dynamical processes in the
system, we derive a single height equation in which coefficients can be related to experimental parameters. This
equation generalizes those obtained by previous continuum models and is able to
account for many experimental features of pattern formation by IBS at oblique
incidence, such as the evolution of the irradiation-induced amorphous layer,
transverse ripple motion with non-uniform velocity, ripple coarsening, onset of
kinetic roughening and other. Additionally, the dynamics of the full two-field
model is compared with that of the effective interface equation.
\end{abstract}


\maketitle


\section{Introduction} \label{introduction}

Materials nanostructuring by ion-beam sputtering (IBS) has received increased
attention in recent years, \cite{valbusa:2002,chan:2007,munoz-garcia:2007} due
to the potential of this bottom-up procedure for applications in
Nanotechnology, and also due to the interesting issues it arises in the wider
context of Pattern Formation at submicrometer scales.\cite{cuerno:2007} In
these experiments, a target is irradiated by a collimated beam of energetic
ions (typical energies being in the keV range) that impinge onto the former
under a well defined angle of incidence. Although routinely employed since long
for many diverse applications within Materials Science (material implantation,
sample preparation, etc.) the capabilities of this technique for efficient
nanopatterning have been recognized only recently, see references in
[\onlinecite{valbusa:2002,chan:2007,munoz-garcia:2007}]. Thus, it induces
self-organized regular ripple (at oblique ion incidence) or dot (at normal ion
incidence, or arbitrary incidence onto rotating targets) nanopatterns over
large areas (up to 1 cm$^2$) on metallic, semiconductor, and insulator surfaces
after a few minutes of irradiation. Interestingly, the main features of this
pattern formation process seem to be largely independent of the type of ions
(even those inducing reactive sputtering) and targets employed, as long as the
latter amorphize under irradiation (the case of metals falls outside this
class, and will not be addressed here, see {\em e.g.} 
[\onlinecite{valbusa:2002,chan:2007,munoz-garcia:2007}]).

During IBS of amorphous or semiconductor substrates (for which the subsurface
layer is amorphized, as frequently observed, see $e.g.$
[\onlinecite{chini:2003,ziberi:2005}]) incident ions loose their energy through
random collisions in the target bulk.\cite{sigmund:1969} Near-surface atoms may
receive enough energy and momentum to break their bonds with the surface. Some
of them may be certainly eroded, but most of them will be redeposited
elsewhere, as seen $e.g.$ in Molecular Dynamics (MD)
simulations.\cite{bringa:2001,moore:2004} In addition to adatom and vacancy
formation,\cite{mayr:2003} which increases surface diffusion currents,\cite{ditchfield:2001}
enhancement of material transport by viscous flow seems to occur within a thin
surface layer, as experimentally verified.\cite{umbach:2001,zhou:2007}
In any case, the evolution of the topography and the appearance of ordered
patterns results from the balance between the erosive and
the relaxational mechanisms. Whereas erosion tends to destabilize the surface
(as a result of the fact that valleys are eroded faster than crests
\cite{sigmund:1973}), relaxational processes tend to reduce height differences.
Although there exists a wide separation of time scales between the hopping
diffusive events (which are of the order of picoseconds) and the ion-impact
events (for an ion flux of $10^{15}$ ions cm$^{-2}$ s$^{-1}$ each atom on a typical surface experiences an ion impact about
once per second), both mechanisms have been modelled using kinetic Monte
Carlo (kMC) approaches. The difference between both scales seems to be
fundamental to correctly describe the evolution of the irradiated surface
---in typical time scales of the order of seconds
\cite{chan:2007,munoz-garcia:2007}--- and challenges description by numerical
simulations.
In order to reach these length scales, a natural procedure is to resort to
continuum descriptions. 
Hence, building upon Sigmund's description of the
(Gaussian) energy distribution for energy deposition from collision cascades
within amorphous or amorphizable targets,\cite{sigmund:1969,sigmund:1973}, the
seminal linear model of Bradley and Harper (BH) \cite{bradley:1988} and its
non-linear extensions \cite{cuerno:1995,park:1999,makeev:2002,kim:2004} already
predict many of the experimentally important features, such as $e.g.$ ripple
formation and orientation as a function of incidence angle and dependencies of
the ripple wavelength with temperature and flux. Moreover, they agree in many
aspects with alternative models, such as kMC studies, see recent discussions in
[\onlinecite{chan:2007}] and [\onlinecite{munoz-garcia:2007}]. 

In all these
continuum models a single evolution equation is formulated for the surface
height field, $h(\mathbf{r},t)$, and contributions to such an equation are
elucidated from the various relaxation mechanisms influencing surface
topography. We will collectively refer to these as one- or single-field models.
Nevertheless, they also present limitations, that we can group into several
categories:

{\em (i) Inaccuracies of the energy distribution:} The fact that there are
known deviations from Sigmund's Gaussian distribution, most conspicuously at
grazing angles of incidence,\cite{nastasi} may account for the incorrect order
of magnitude of the roughening rate as estimated by these models, or their
incorrect prediction \cite{alkemade:2006} for the direction of transverse ripple motion.

{\em (ii) Restricted number of mechanisms:} Continuum models
necessarily neglect physical mechanisms which may turn out to be important to
the system behavior. This fact may be related with the unsatisfactory
description by one-field models of the ripple wavelength dependence with
energy, phenomena such as pattern wavelength coarsening or, for the case of
normal ion incidence, their lack of account for in-plane ordering, or the high parameter sensitivity for dot formation. 

{\em (iii) Formal consistency:} Under some circumstances,
the very formal consistency of the one-field models can be questioned. For
instance, due to the {\em ad-hoc} nature of the way in which
competing physical effects (such as physical sputtering and surface diffusion)
are merely {\em added} in the height equation of motion. Or due to the
existence of cancellation modes of a varying nature
\cite{rost:1995,kim:2004,castro:2005b,kim:2005} in the non-linear equations, or
to physically unstable values of the effective surface diffusion coefficients
for intermediate incidence angles.\cite{makeev:2002} 

{\em (iv) Non-linear
features:} Finally, the explanation for some of the experimental properties
that remain insufficiently accounted for by previous continuum models may
require improvements on our understanding of non-linear effects (and thus,
affect any further continuum descriptions). Some of these may include the direction of transverse ripple
motion, the spread in the measured values of roughness exponents when there is
kinetic roughening, and the value (as a function of physical parameters) of the
saturated ripple or dot amplitude.

Due to the insufficiencies of the current continuum descriptions of pattern
formation by IBS, we conclude on the need for improved continuum models that
{\em (a)} introduce increased number and/or type of relaxation mechanisms in a
natural way, that in particular allow assessment of the interplay between
transport and morphology; {\em (b)} improve upon consistency issues
(cancellation modes, etc.) of previous approaches; {\em (c)} can be adapted to
modifications in the distribution of energy deposition; {\em (d)} can account
for the phenomenology of nanopatterning by IBS within an unified framework, and
{\em (e)} generalize previous linear and non-linear models, incorporating their
successes and improving upon their shortcomings.

Trying to reach a balance between complexity and completeness in the physical
description, in [\onlinecite{aste:2004,aste:2005,castro:2005,munoz-garcia:2006}] continuum models have been considered that are simpler than a full hydrodynamic
description but still provide an explicit coupling between the surface
topography and the evolution of the relevant diffusive fields. Following the
philosophy behind the so-called ``hydrodynamic'' approach to aeolian sand
dunes, \cite{terzidis:1998,valance:1999,csahok:2000} in order to describe the
temporal evolution of the topography, two coupled fields are considered,
namely, the density of mobile species being transported at the surface and the
local height of the static target. Although naturally there are important
differences between IBS nanopatterns and ripples on aeolian sand dunes ($e.g.$
in IBS the size of the structures is roughly seven orders of magnitude smaller,
the total mass is not conserved due to sputtering, and the nature of the
morphological instability resides in the erosive process, rather than in the
transport processes, as a difference with wind of water induced patterns on
granular systems) both of them share global features that suggest modeling
along similar lines.

In this paper we study in detail this two-field approach to IBS, expanding
previous results obtained in Refs.\
[\onlinecite{castro:2005,munoz-garcia:2006}], and focusing on the most generic
case of arbitrary (oblique) angle of incidence that is pertinent to ripple
formation. We will assess the extent to which two-field models can contribute
to the improvement of continuum description of IBS as described in points $(a)$
through $(e)$ above, and can be seen as a continuum reformulation of thin film
surface dynamics that goes even beyond the specific instance of IBS. 
Our aim here is also to clarify the influence of different experimental
parameters, such as temperature or ion flux energy, in order to stimulate
new controlled experiments. We derive
an improved interface equation and relate the parameters appearing in it to
experimental conditions and features. In a companion paper,
\cite{munoz-garcia:2007b} that will be henceforth denoted as paper II, we explore the
implications of our two-field model for the cases of normal ion incidence and
rotating targets, that are of interest $e.g.$ for the production of quantum dots
by this experimental technique.

This paper is organized as follows. In the next section
the basic ideas of the coupled two-field model are discussed. In section
\ref{linear} its planar solution is obtained and a linear stability
analysis is performed. Section \ref{nonlinear} is devoted to obtaining a single effective evolution equation to describe the surface height of the bombarded surface, by means of a
multiple scale analysis. In order to check the hypothesis made in the
derivation of that effective equation, in Sec.\ \ref{1-D_model} the dynamics of this equation will be compared with that of the original two-field model in the
1D case. Following this, we will study the two dimensional interface equation in Sec.\ \ref{2-D_equation}, and consider its relation to experiments. To end, we provide our main conclusions in Sec.\ \ref{Conclusions}. In the appendices we collect details of several analytical calculations.

\section{Model} \label{model}

For the model formulation, a key experimental fact for amorphous and
amorphizable targets is the formation through irradiation of a thin amorphous
layer at the target surface, see references in
[\onlinecite{chan:2007,munoz-garcia:2007}]. As done in Refs.\
[\onlinecite{aste:2004,aste:2005,castro:2005,munoz-garcia:2006}], the main
model assumption is that the surface dynamics can be completely described
through the time evolution of two fields: the height $h(\mathbf{r},t)$ of the
static substrate at time $t$ and point $\mathbf{r}=(x,y)$ on a reference plane that
coincides with the uneroded flat surface, and the thickness $R(\mathbf{r},t)$
of the (thin) surface layer of mobile species. This thickness can be related with the density of, say,
mobile adatoms through their atomic volume. Note that for the energies we are
considering in the order of 1 keV, we can take adatoms as the dominant
diffusing species, although $e.g.$ for energies below 1 keV, advacancies may
dominate surface transport effects;\cite{chan:2007} this should reflect in the
values of the diffusion constant to be introduced below.

Dynamics of the two fields are coupled, and read
\begin{eqnarray}
\partial_t R &=& (1-\phi) \Ge - \Ga - \nabla \cdot \mathbf{J}, \label{eq.R}\\
\partial_t h &=& -\Ge+\Ga, \label{eq.h}
\end{eqnarray}
where the $\mathbf{\hat{x}}$ axis is chosen as the projection of the beam
direction onto the $xy$ plane. In (\ref{eq.R})-(\ref{eq.h}), $\Ge$ and $\Ga$,
which depend on the geometry of $R$ and $h$, are, respectively, the rate at
which material is dislodged from the immobile target due to irradiation (locally
decreasing the value of $h$), and the rate at which mobile material
incorporates back into the immobile bulk (locally increasing the value of $h$).
Therefore, in opposition to the excavation mechanism which is responsible for
the overall decrease of $h$, there exists a process of incorporation back to
the bulk analogous of a local condensation of mobile species. Nevertheless, we
will {\em not} consider a spontaneous rate of ``evaporation'' that is
independent of the ion beam, so that we are neglecting surface tension-mediated
evaporation/condensation effects \cite{mullins:1957,mullins:1959}
(equivalently, we are assuming that the pressure in the vapor phase is
negligible). The excavated material may be either sputtered away, or added back
to the mobile thickness $R$ with an efficiency $(1-\phi) \equiv \bar{\phi}$.
Therefore, the fraction of the eroded atoms which are finally sputtered away is
represented by $\phi$ so that, for ${\phi} \neq 1$, local redeposition is
partially allowed.\cite{kustner:1998} For ${\phi}=1$ all eroded atoms are
sputtered away, while in the $\phi=0$ case the sputtering yield is zero. In the
last case the effect of the ion beam is limited to providing material for
surface transport, and there is no average motion of the interface. We will
refer to the latter two cases as zero-redeposition and complete redeposition
limits, respectively. They will constitute useful limiting cases below. 

The system (\ref{eq.R})-(\ref{eq.h}) was put forward in Refs.\
[\onlinecite{aste:2004,aste:2005}], in which a
linear stability analysis was performed.
However, one of the limitations of the choices made in these works for $\Ge$
and $\Ga$ is that surface diffusion vanishes in the absence of redeposition
${\phi}=1$, making the ensuing model ill defined (due to a large wave-vector
instability). These limitations were overcome in Refs.\
[\onlinecite{castro:2005,munoz-garcia:2006}], in which more physical mechanisms of
erosion and addition are considered.

The third term on the right hand side of Eq.\ (\ref{eq.R}) describes transport of
mobile material onto the surface in the form of a continuity equation. In
contrast to [\onlinecite{aste:2004,aste:2005}], where terms representing
Erlich-Schwoebel barrier effects (relevant to IBS of metals
\cite{valbusa:2002}) are incorporated into the diffusive current $\mathbf{J}$,
these are not considered in [\onlinecite{castro:2005,munoz-garcia:2006}]. With
the aim of studying amorphous or semiconducting targets we will follow the
latter option. Here we simply consider a diffusive term for mass transport onto
the surface that, in the case of isotropic materials, is given by
$\mathbf{J}=-D \nabla R$, where $D$ may be a temperature dependent constant (see
below). 

Likewise, we will neglect momentum transfer in the direction of the
projection of the beam of ions to adatoms, as this is expected to be
non-negligible only at higher energies (say, above $10^3$ keV, see $e.g.$
[\onlinecite{cliche:1995}] and a discussion in [\onlinecite{chan:2007}]).

\subsection{Excavation}

We next need expressions for the excavation and addition rates. As studied in
previous theoretical single-field
studies,\cite{bradley:1988,cuerno:1995,makeev:2002} the rate at which material
is sputtered from the bulk depends on experimental conditions such as the angle
of incidence, $\theta$, substrate and ion species, ion flux, $\Phi$, average
ion energy, $E$, temperature, $T$, and other. In these works, such dependencies
were studied through an assumption on the shape of the spatial distribution for
energy deposition, mostly Sigmund's Gaussian distribution. However, there are
cases in which systematic deviations from the Gaussian shape occur (see $e.g.$
[\onlinecite{feix:2005}] for the occurrence of {\em exponential} decay combined
with null energy deposition along the ion track). As recently shown
moreover,\cite{davidovitch:2007} the shape of this distribution may affect the
very existence of a morphological instability and thus the formation of a
pattern. At any rate, given the fact that for most ripple patterns the aspect
ratio is small enough so that a small slope approximation is expected to hold
\cite{chan:2007} (except, possibly, for compound materials and predesigned
substrates \cite{chen:2005}), to lowest non-linear order, the {\em form} of the
excavation rate is expected to be \cite{castro:2005,munoz-garcia:2006}
\begin{multline}\label{Ge}
  \Gamma_{ex}  =  \alpha_0 [ 1 + \alpha_{1x}\, \partial_x h + \nabla \cdot (\underline{\alpha_2} \nabla h)+ \partial_x \nabla \cdot (\underline{\alpha_3} \nabla h) \\
  + \nabla \cdot (\underline{\alpha_4} \nabla \nabla^2 h) + \partial_x h \, \nabla \cdot (\underline{\alpha_5} \nabla h) + \nabla h \cdot (\underline{\alpha_6} \, \nabla h) ]
\end{multline}
independently of the assumed energy distribution.
\cite{bradley:1988,cuerno:1995,makeev:2002,feix:2005} 
Here, we will ignore the effects of direct erosion (knock-on sputtering) which could be relevant under very shallow energy deposition conditions ($e.g.$ at very grazing angles of incidence). Indeed, the local erosion
velocity that follows from Sigmund's distribution has the shape given in
(\ref{Ge}), see [\onlinecite{makeev:2002}] and Appendix \ref{app.A}. Changes in
the energy distribution are of course expected to modify the {\em values} of
the parameters, but not the number and shape of the terms appearing in
(\ref{Ge}), that are a consequence of the loss of $x \leftrightarrow -x$
symmetry induced by the oblique beam. Note that reflection symmetry is not lost
in the $y$ direction, and that the $x-y$ symmetry can be restored under
different incidence conditions, such as normal incidence ($\theta=0$) and for
rotating targets, see paper II.
Thus, we have that in general $\underline{\alpha_i}={\rm
diag}(\alpha_{ix},\alpha_{iy})$ are $2\times 2$ diagonal matrices for
$i=2,3,5,6$, while $\underline{\alpha_4}= \left[\begin{array}{cc} \alpha_{4xx}
& \alpha_{4xy} \\ \alpha_{4yx} & \alpha_{4yy} \end{array} \right]$.

The parameter $\alpha_0$ defines the excavation rate of a
flat surface and is directly related to the sputtering yield of a flat surface,
$Y_0$, the ion flux, $\Phi$, and the number of atoms per unit
volume in the solid, $n_v$, by $\alpha_0=\Phi Y_0/n_v$. Since typical fluxes
range from $\Phi=10^{12}$ cm$^{-2}$ s$^{-1}$ to $\Phi=10^{17}$ ions cm$^{-2}$
s$^{-1}$, the number of atoms per unit volume for an atomic diameter of
$0.4$ nm is $n_v = 30$ nm$^{-3}$, and typical yields for experiments with ion
energies of some keV are of order unity, then $\alpha_0 \thickapprox
10^{-3}-10^2$ nm s$^{-1}$.

While the detailed dependence of the remaining $\alpha_{ijk}$ coefficients on the physical parameters can be rather non-trivial, the main physical content of
Eq.\ \eqref{Ge} is relatively straightforward. Thus, $e.g.$, as already shown by
BH, the coefficients $\alpha_{2x}$ and $\alpha_{2y}$ are positive (see Appendix
\ref{app.A}) at small angles of incidence, which implies faster excavation at
surface minima than at surface maxima, which is the landmark of Sigmund's
morphological instability. Similarly, the various terms in \eqref{Ge} imply
{\em geometrical} dependencies of the excavation rate with surface morphology;
say,\cite{makeev:2002} for small $\theta$ one has  $\alpha_{1x} < 0$ so that the excavation
rate is larger on a lee ($\partial_x h < 0$) ripple slope than on a stoss
($\partial_x h > 0$) ripple slope. However, we will see that, when coupled to
surface transport, some of these dependencies can be modified with respect to
the simplest expectations. Conspicuous geometrical dependencies of this sort
appear through the coefficients of the $\underline{\alpha_4}$ tensor. Within
Sigmund's energy distribution, these are high order geometrical dependencies of
the sputtering rate that in one-field equations reflect into terms with the
shape of surface diffusion. However, the present formulation makes it
transparent the extent to which such terms do {\em not} correspond to actual
material transport on the surface. We will come back to this point later.

\subsection{Addition} \label{Sub_addition}

One-field models are basically complete once $\Ge$ is provided. However, in our
case we still need to specify the addition rate $\Ga$. To this end, we have to
take into account that surface diffusion is an independent physical mechanism
that can take place even in the absence of an ion beam. Of course it should be
susceptible of enhancement by the presence of the latter due to the induced
increase in the density of diffusing species, but within our framework we would
like to have surface diffusion currents which are not necessarily proportional to the ion flux. To this
end, we will allow for a non-zero thickness of mobile material $R_{eq}$ even in
the absence of excavation ($\Gamma_{ex}=0$) or redeposition ($\phi=0$),
and write down a rate that favors addition in highly coordinated surface
positions (minima) rather than at sites with low coordination (surface maxima).
Thus, we write \cite{castro:2005}
\begin{equation}\label{Ga}
\Gamma_{ad}=\gamma_0\left[R-R_{eq}(1-\gamma_{2x}\partial^2_xh-\gamma_{2y}
\partial^2_yh)\right],
\end{equation}
that has a form that is reminiscent from the Gibbs-Thompson expression effect
for surface relaxation via
evaporation-condensation.\cite{mullins:1957,mullins:1959} In Eq.\ \eqref{Ga},
$\gamma_0$ is the mean nucleation rate for a flat surface, $\gamma_0^{-1}$
representing the typical time between two nucleation events, typically in the
range of picoseconds, and $\gamma_{2x}$, $\gamma_{2y} \geq 0$ describe the
variation of the nucleation rate with the surface curvatures. In principle
this paper focuses on amorphous or amorphizable surfaces, for which $\gamma_{2x} = \gamma_{2y} \equiv \gamma_2$ although, for the sake of generality, we will consider the most general case of anisotropic nucleation rates ($\gamma_{2x} \neq \gamma_{2y}$) as far as convenient. 

As we will see later, the thickness of the mobile material, $R$, is only slightly altered off its equilibrium value, $R_{eq}$, so that the rate of addition previously considered in [\onlinecite{munoz-garcia:2006}] is equivalent to \eqref{Ga}, at least sufficiently close to the instability threshold. We will
see in the next section that \eqref{Ga} indeed leads to proper surface
diffusion effects, that will allow us to identify the phenomenological
parameters $R_{eq}$, $D$ and $\gamma_2$ with physical constants.


\section{Planar solution and linear stability analysis} \label{linear}

The existence of a wide separation of time scales between diffusive events and
erosive events will allow us to simplify the study of model
\eqref{eq.R}-\eqref{Ga}. We can assume than the excavation rate, $\alpha_0$, is
much smaller than any other velocity involved in the problem. Specifically, by
considering $\alpha_0 \ll \gamma_0 R_{eq}$, we can define a non-dimensional
parameter $\epsilon=\alpha_0/(\gamma_0 R_{eq})$ which will simplify the study of
the system in the following sections. As noted above, typically $\alpha_0
\thickapprox 10^{-3}-10^{2}$ nm s$^{-1}$, while the frequency of hopping
diffusive events, equivalent to $\gamma_0$, is of the order
\cite{dichtfield:2001} of $10^9$ s$^{-1}$. If we consider that the thickness of
the mobile layer in equilibrium is of the order of some atomic sizes,
$R_{eq}\thickapprox 1$ nm, we get as an estimate for typical values of
$\epsilon$ to be in the range $\epsilon \thickapprox 10^{-12}-10^{-7}$, larger
values corresponding to higher fluxes and/or larger yield conditions.

\subsection{Planar solution} \label{Sub_planar}

In order to start the study of our model, we first consider the situation of a
perfectly flat interface. In such a case, all the spatial derivatives of
$h(\mathbf{r},t)$ are zero, Eqs.\ \eqref{eq.R} and \eqref{eq.h} becoming
\begin{align}
  \partial_t R^p &= \epsilon \bar{\phi}
\gamma_0 R_{eq}-\gamma_0 \left( R^p-R_{eq} \right),\label{eq.Rp}\\
    \partial_t h^p &= -\epsilon \gamma_0 R_{eq}+\gamma_0 \left( R^p-R_{eq} \right) \label{eq.hp},
\end{align}
where we have defined $R^p(t)$ and $h^p(t)$ as the planar solution fields.
Integrating Eq.\ \eqref{eq.Rp} and assuming $R(t=0)=R_{eq}$ we obtain $R^p$,
which reads
\begin{equation}\label{Rp}
    R^p(t)=R_{eq} \left[1 + \epsilon \bar{\phi} (1-e^{-\gamma_0 t})\right],
\end{equation}
for any value (not necessarily small) of $\epsilon$. In \eqref{Rp}
we see that, after a short time (of the order of $\gamma_0^{-1}$), $R^p$ reaches
a stationary value equal to $R_{eq}$, plus a small modification of order
$\epsilon$ due to the redeposition of excavated material (such an extra term is
absent in the zero redeposition, $\phi=1$, case). As indicated in Sec.\ \ref{Sub_addition}, even in the
absence of excavation ($\alpha_0=\epsilon=0$) or redeposition ($\bar{\phi}=0$),
there still exists an intrinsic fraction of mobile material equal to $R_{eq}$.

Substituting \eqref{Rp} into \eqref{eq.hp} and assuming that $h(t=0)=0$, we
obtain the evolution of the planar height of the bombarded surface, namely,
\begin{equation}\label{hp}
h^p(t) = \epsilon R_{eq} \left[- \phi \gamma_0 t +\bar{\phi}\left(e^{-\gamma_0 t}-1\right) \right] \to -v_0 t ,
\end{equation}
where the last expression holds for times longer than $\gamma_0^{-1}$, for which the planar profile {\em erodes} with a constant velocity $v_0 =
\epsilon \phi \gamma_0 R_{eq}=\phi\alpha_0$. This expression gives
a clear interpretation of the parameter $\phi$ as the overall efficiency of the
sputtering process.

\subsection{Linear stability analysis}\label{Linear_stability_analysis}

The next step is to perform a linear stability analysis in order to investigate
whether a small perturbation of the planar solution is amplified or damped out
in the course of time. We consider periodic perturbations of the form
\begin{align}
    R(\mathbf{x},t)&=R^p(t)+{R_0^\ell} \exp(i\mathbf{k \cdot r} +\omega_{\mathbf{k}} t),\label{Rl} \\
    h(\mathbf{x},t)&=h^p(t)+{h_0^\ell} \exp(i\mathbf{k \cdot r} +\omega_{\mathbf{k}} t)\label{hl},
\end{align}
where $\mathbf{k}=(k_x,k_y)$ is the wave vector of the perturbation and
$\omega_{\mathbf{k}}$ its dispersion relation. Substituting Eqs.\ \eqref{Rl}
and \eqref{hl} into \eqref{eq.R} and \eqref{eq.h}, and neglecting quadratic
terms in $R_0^\ell, h_0^\ell$, we obtain the following linear system of equations
\begin{equation} \label{sistemalineal}
 \left(\begin{array}{cc} \omega_{\mathbf{k}} +\gamma_0 + D\mathbf{k}^2 &
-\bar{\phi}\Gamma^\ell_{ex} + \Gamma^\ell_{ad}        \\
  -\gamma_0 & \omega_{\mathbf{k}}+\Gamma^\ell_{ex}-\Gamma^\ell_{ad}
   \end{array}\right)
    \left( \begin{array}{cc} {R}_0^\ell \\ {h}_0^\ell \end{array}
\right) = 0,
\end{equation}
where
\begin{align}
    \Gamma^\ell_{ex} =& \epsilon \gamma_0 R_{eq} \left[\alpha_{1x}i k_x - \sum_{j=x,y}\left(
    \alpha_{2j} + \alpha_{3j} i k_x \right) k_j^2 \right.\nonumber \\
    & - \left. \sum_{i=x,y} \alpha_{4ij} k_i^2 k_j^2 \right], \label{Gelineal}\\
    \Gamma^\ell_{ad} =& -\gamma_0 R_{eq} (\gamma_{2x} k_x^2+\gamma_{2y} k_y^2) \label{Gadlineal}.
\end{align}
Non-trivial solutions only exist when the determinant of the coefficient matrix
equals zero, which allows us to obtain the dispersion relation
$\omega_{\mathbf{k}}$ as the solution of the following complex second order
equation
\begin{equation}\label{omega}
    \omega_{\mathbf{k}}^2+\omega_{\mathbf{k}} (a+ib)+(c+id)=0 ,
\end{equation}
where the coefficients $a$, $b$, $c$, and $d$ are functions of parameters and
wave-vector components, and are given in Appendix \ref{app.AA}. Eq.\
\eqref{omega} leads to two branches in the dispersion relation, corresponding
to its two (complex) solutions, namely,
\begin{align}
    {\rm Re}(\omega_{\mathbf{k}}^\pm) & =  -\frac{a}{2} \pm
    \frac{1}{2\sqrt{2}} \left(\left[(a^2-b^2-4c)^2+(2ab-4d)^2\right]^{1/2} \right.
    \nonumber \\
    &  \left. +a^2-b^2-4c \right)^{1/2} ,\label{ReW+-}\\
    {\rm Im}(\omega_{\mathbf{k}}^\pm) & =  -\frac{b}{2} \pm \frac{1}{2\sqrt{2}}
    \left(\left[(a^2-b^2-4c)^2+(2ab-4d)^2\right]^{1/2} \right. \nonumber \\
    &  \left. -a^2+b^2+4c \right)^{1/2} \label{ImW+-}.
\end{align}
Substituting Eqs.\ \eqref{ap_a}-\eqref{ap_d} for $a$, $b$, $c$, and $d$ into Eqs.\ \eqref{ReW+-} and \eqref{ImW+-},
we obtain an analytical expression for the dispersion relation as a function of
the model parameters. Thus, we can describe the linear evolution of a periodic
perturbation to the planar solution, since the real part of
$\omega_{\mathbf{k}}$ is related to the growth or decay of the perturbation
amplitude, while the imaginary part describes its in-plane propagation. Since
we are interested in the behavior of the system for long distances, we will
reduce our analysis of $\omega_{\mathbf{k}}$ to small wave vectors. In this
limit, we get, to lowest order in $k_x$ and $k_y$,
\begin{align}
    {\rm Re}&(\omega_{\mathbf{k}}^-)=-\gamma_0,\label{ReW-}\\
    {\rm Re}&(\omega_{\mathbf{k}}^+)= \epsilon \phi \gamma_0 R_{eq} \left(
    \alpha_{2x}k_x^2+\alpha_{2y}k_y^2- \epsilon \bar{\phi} \alpha_{1x} k_x^2 \right).\label{ReW+}
\end{align}
Thus, the negative branch is unconditionally stable (perturbations decay
exponentially for any wave vector) and non-trivial dynamics (including the
pattern formation process) are thus governed by the positive branch, which
features a band of unstable modes (wave vectors), of small magnitude for small
$\epsilon$ values, for which perturbations can grow exponentially. The
imaginary part of the dispersion relation for $k\equiv  |\mathbf{k}| \ll1$ is ${\rm
Im}(\omega_{\mathbf{k}})=-\epsilon \phi \gamma_0 R_{eq} \alpha_{1x} k_x$, or
${\rm Im}(\omega_{\mathbf{k}})=- \epsilon \bar{\phi} \gamma_0 R_{eq}\alpha_{1x}
k_x$ depending of the branch, the sign of $k_x$, and the sign of $(\phi-1/2)$. The
linear in-plane propagation of the perturbations is related to the imaginary part of the dispersion relation [see Eq.\ \eqref{v1d_b} below]. For the positive branch, positive modes and large
redeposition ($\phi<1/2$), we have
\begin{equation} \label{ImW}
{\rm Im}(\omega_{\mathbf{k}}^+)=- \epsilon \phi \gamma_0 R_{eq}\alpha_{1x} k_x,
\end{equation}
which indicates that the perturbations travel along the $x$ direction with a
constant velocity equal to $\alpha_0 \phi \alpha_{1x}$.

With respect to the time evolution of the amplitude of perturbations, the
linear pattern features are provided by that mode which makes ${\rm
Re}(\omega_{\mathbf{k}}^+)$ a positive maximum. For, say, small angles of
incidence, both $\alpha_{2x}$ and $\alpha_{2y}$ are positive \cite{makeev:2002}
so that \eqref{ReW+} is maximized for infinite wave vector components. A {\em
finite} maximum is seen to occur once we take into account higher order
corrections (in $k$) to \eqref{ReW+}, where stabilizing mechanisms compete with
the erosion instability. Thus,
\begin{multline}\label{ReW}
 {\rm Re} (\omega_{\mathbf{k}}^+)  =  \epsilon \phi \gamma_0 R_{eq}(\alpha_{2x}k_x^2+
    \alpha_{2y}k_y^2-\epsilon\bar{\phi}\alpha_{1x}k_x^2)  \\
      -  R_{eq}Dk^2(\gamma_{2x}k_x^2+\gamma_{2y}k_y^2) \\
     - \epsilon \gamma_0 R_{eq} \sum_{i,j=x,y} \Big[\phi\alpha_{4ij}
       - \left. \left(\frac{ \bar{\phi} D}{\gamma_0} - \phi R_{eq} \gamma_{2i} \right) \alpha_{2j}\right]
     k^2_i k^2_j ,
\end{multline}
where we have kept terms that are lower order than $O(\epsilon^2k^4)$. Eq.\
\eqref{ReW} has the same form as the corresponding expression in one-field
theories but with modified coefficients, see Appendix \ref{app.B}. In general,
the $O(k^2)$ terms are both of a purely erosive origin, being directly
proportional to the curvature dependencies of the excavation rate (once we
neglect the $O(\epsilon^2)$ contribution), that are available for several
energy distribution functions.\cite{bradley:1988, makeev:2002, feix:2005} Thus,
in particular, our model respects the signs of these terms as obtained $e.g.$ by
BH.\cite{bradley:1988} Given their destabilizing nature, they
are usually referred to as ``negative'' surface tension terms. The remaining
$O(k^4)$ terms in \eqref{ReW} are of an opposite stabilizing nature related to
surface diffusion effects as justified below.


\subsubsection{Two-field description of surface diffusion} \label{sec.Mullins}

In order to clarify the physical meaning of the $O(k^4)$ contributions in
\eqref{ReW}, it is useful to consider different relaxation mechanisms that are
known to lead to such type of terms.

\paragraph{Thermal surface diffusion} \label{thermal_surface_diffusion}

Let us study at this point the extreme limit of no erosion in the original
model \eqref{eq.R}-\eqref{Ga}. This can be achieved by simply ``turning off''
the ion beam flux setting $\alpha_0=0$, which in turn implies $\epsilon=0$.
Note that, physically, in this case we are left with a system in which
variations in the substrate height $h$ are only due to local
detachment/addition and transport of the surface mobile species $R$, precisely
as in Mullins' classic description of surface diffusion activated by
temperature.\cite{mullins:1957,mullins:1959} Mathematically, the dynamics of
the ensuing system \eqref{eq.R}-\eqref{eq.h} conserves the total amount of
material and, moreover, dynamics are linear (note, nonlinearities enter only
through the rate $\Ge$, that has been turned off). Thus, one can readily solve
the full system in this case. To our purposes we are interested in the long
wavelength limit, for which we can simply take the $\epsilon \to 0$ limit in
the results of the present Section. Up to order $O(k^4)$, and already
restricting ourselves to the isotropic case
$\gamma_{2x}=\gamma_{2y}=\gamma_2$, the result is
\begin{align}
{\rm Re} (\omega_{\mathbf{k}}^+) & =  - R_{eq} D \gamma_2 k^4, \\
{\rm Im} (\omega_{\mathbf{k}}^+) & =  0 .
\end{align}
Thus, the exact evolution equation for the surface height in this
long-wavelength limit reads
\begin{equation}
\partial_t h = -R_{eq} D \gamma_2 \nabla^4 h , \label{eq.sd}
\end{equation}
to be compared with Mullins' result,\cite{mullins:1957,mullins:1959} namely,
\begin{equation} \label{eq.Mullins}
\partial_t h = - \frac{D_s \nu_s \gamma}{k_B T n_v^2} \nabla^4 h ,
\end{equation}
where $D_s$ is the surface diffusivity of mobile surface species, $\nu_s$ is
their concentration, $\gamma$ is the surface free energy per area, $k_B$ is
Boltzmann's constant, and $n_v^{-1}$ is the atomic volume. From this we see
that the corresponding contribution in \eqref{ReW} is a generalization of
surface diffusion in which the surface free energy is taken to be anisotropic
in the two substrate directions. Moreover, with the use of dimensional
arguments, we identify parameters in $\Ga$ as $\gamma_2 = \gamma/(k_B T n_v)$,
$D = D_s$, and $R_{eq} = \nu_s/n_v$,
whereby $\Ga$ becomes an implementation of Gibbs-Thompson
formula.\cite{mullins:1959} In any case, we see that the applicability of the
two-field approach goes beyond the specific case of erosion by IBS, and it can
serve as an intuitive phenomenological reformulation of other phenomena within
Surface Science.

\paragraph{Surface confined viscous flow}

It is also a classic result \cite{orchard:1962} that viscous flow, when
confined to a thin surface layer, leads to a contribution to the height
evolution of a similar form to \eqref{eq.Mullins}
\begin{equation}
\partial_t h = - \frac{d^3 \gamma}{\eta_s} \nabla^4 h , \label{eq.Orchard}
\end{equation}
where $d$ is the thickness of the viscous layer and $\eta_s$ is the viscosity.
In the case of IBS erosion of silicon targets, the relevance of such type of
relaxation mechanism has been pointed out.\cite{umbach:2001} Specifically, it
is argued in [\onlinecite{umbach:2001}] that the ion beam induces this type of
flow in such a way that $1/\eta_s \propto E \Phi$, where $E$ is the average ion
energy. Notice that, under this assumption, all $O(k^4)$ terms in \eqref{ReW}
would become proportional to ion energy and flux. 

In general, one expects both effects, thermal surface diffusion and ion-induced
surface viscous flow, to occur simultaneously in IBS systems,\cite{ditchfield:2001} so that an equation like \eqref{eq.sd} should account for the effects described by \eqref{eq.Mullins} and \eqref{eq.Orchard}.
A form to accommodate this fact is to assume on a phenomenological basis that $R_{eq}$ and $D$ include both thermal (i.e.\ beam independent) and beam dependent contributions.


\subsubsection{Features of the linear instability} \label{linear_inst}

We now come back to the full IBS model ($i.e.$, for generic $\alpha_0 \neq 0$).
Note that there are up to three different $O(k^4)$ terms [second and third line in Eq.\ \eqref{ReW}]. Besides thermal
surface-diffusion of the type discussed in Sec.\ \ref{sec.Mullins}, the terms
proportional to $\alpha_{4ij}$ [on the last line of Eq.\ \eqref{ReW}] originate in the high order dependence of the
excavation rate $\Ge$ with the height derivatives, and correspond to the
so-called ``effective smoothing'' terms in one-field
models.\cite{makeev:1997,makeev:2002} As is clear from our present formulation,
being independent of $R_{eq}$ and $D$, these terms do {\em not} originate in
actual material transport on the surface.\cite{footnote2} In marked contrast,
the remaining $O(k^4)$ terms on Eq.\ \eqref{ReW} do couple excavation (they are
proportional to $\alpha_{2i}$) to surface transport (being proportional to
either $D$ or $R_{eq} \gamma_{2i}$), a feature that is beyond one-field
descriptions. In particular, they may become temperature-dependent through the
latter parameters, which will have relevant implications below.
Similarly to one-field models, ``surface diffusion" like terms oppose the
erosive instability and lead to selection of a typical length-scale in terms of
the wave vector which grows (linearly) fastest. From \eqref{ReW}, we can obtain
the features (orientation and magnitude) of such mode providing the ripple
structure.

\paragraph{Ripple orientation}

Using the results in Appendix \ref{app.B}, for the small physically relevant
values of $\epsilon$, the ripple structure can only align along the $x$ or the
$y$ directions.
Using \eqref{ReW}, for isotropic thermal surface diffusion,
$\gamma_{2x}=\gamma_{2y}$, the ripple pattern is oriented along the $x$
direction (with crests aligned in the $y$ direction) if $\alpha_{2x} >
\alpha_{2y}$, or in the $y$ direction (with crests aligned in the $x$ direction)
when $\alpha_{2y} > \alpha_{2x}$, or is a linear superposition of the two
orientations when $\alpha_{2y} = \alpha_{2x}$, in which case one has a
square-symmetric cell arrangement, rather than a proper ripple structure. These
results for the ripple orientation generalize those of one-field
models,\cite{bradley:1988,makeev:2002} for which there is moreover abundant
experimental confirmation, see references $e.g.$ in
[\onlinecite{chan:2007,munoz-garcia:2007}]. When thermal surface diffusion is
anisotropic, $\gamma_{2x} \neq \gamma_{2y}$, the possibilities of alignment for
the ripple pattern are again along the $x$ axis, along the $y$ axis, or
simultaneously in both directions (corresponding to an array of rectangular
cells) if $\alpha_{2x}^2 \gamma_{2y}=\alpha_{2y}^2 \gamma_{2x}$.

\paragraph{Ripple wavelength}

In the cases above, the leading contribution (in powers of $\epsilon$) of the
wave vector at which the linear dispersion relation is maximized reads
\begin{equation}\label{kl}
k^{\ell}_{x,y}\simeq\sqrt{\frac{\epsilon \phi \gamma_0 \alpha_{2x,y}}{2D\gamma_{2x,y}}} =
\sqrt{\frac{\alpha_0 \phi \alpha_{2x,y}}{2R_{eq}D\gamma_{2x,y}}} ,
\end{equation}
where the $x$ (resp.\ $y$) subindex applies when the ripples align in the $x$
(resp.\ $y$) direction. Recalling the order of magnitude of the model
parameters as given in the previous section, we can substitute them into
\eqref{kl}. Assuming further $\alpha_{2x,y}$ and $\gamma_{2x,y}$ to be of the
same order of magnitude ($e.g.$, assuming that the only relaxational mechanism
is thermal surface diffusion and employing the relations given in Sec.\ 
\ref{thermal_surface_diffusion}), we have $\alpha_{2x,y} = 0.18$ nm and
$\gamma_2= 3.8$ nm using data for Si(001) as in [\onlinecite{erlebacher:2000}]
for $T=500^{\circ}$C, and $D \approx 10^5$ nm$^2$ s$^{-1}$ as measured in
[\onlinecite{dichtfield:2001}]. We thus obtain $k^{\ell} \approx \left(10^{-4} -
10^{-2}\right)R_{eq}^{-1/2}$ nm$^{-1}$, where we have used values for $\alpha_0
= 10^{-3}-10^2$ nm s$^{-1}$ as above and the thickness of the mobile surface
species layer, $R_{eq}$, must be given in nm. If we consider this thickness to
be comparable to a few atomic diameters, $R_{eq}\approx 1$ nm, we finally obtain
an estimate of the linear ripple wavelength ${l}^\ell= 2\pi/k^{\ell}$. Thus, ${l}^\ell \approx 10-10^4$ nm, in agreement with the experimental
orders of magnitude.\cite{munoz-garcia:2007}

Subdominant contributions to the ripple wavelength are physically very informative of the interplay among the physical mechanisms present in the two-field model. Thus, for instance in the case of ripples along the $x$ direction one gets to next order in $\epsilon$
\begin{equation} 
{l}^\ell = 
2^{3/2}\pi \left( \frac{D R_{eq}
\gamma_{2x}} {\phi\alpha_0 \alpha_{2x}} - \frac{\Delta_x}{\phi}  +
\frac{\alpha_{4xx}}{\alpha_{2x}} \right)^{1/2} \label{l1d},
\end{equation}
where we have used the parameter combinations
\begin{equation}
\Delta_{i} = \frac{\bar{\phi}D}{\gamma_0}-\phi R_{eq} \gamma_{2i} , \quad i=x,y .\label{eq.Deltas}
\end{equation}
In view of the physical interpretation of the various parameters entering Eq.\
\eqref{l1d}, we see that the argument of the square root in
this expression is the sum of a temperature independent contribution (the term $\alpha_{4xx}/\alpha_{2x}$) corresponding to the ion-induced effective diffusion of Ref.\ [\onlinecite{makeev:2002}] and terms which include both thermal and beam dependent contributions. Such a compound structure for the linear ripple wavelength coincides precisely with that employed by Umbach et al.\cite{umbach:2001} when showing the importance of surface viscous flow in order to account for the experimental behavior of the ripple wavelength with flux and temperature. It also has the same shape as that proposed in
[\onlinecite{chan:2007}], capturing in a phenomenological way various
experimental observations. We again stress that formula \eqref{l1d} is
obtained within a linear approximation for which the ripple wavelength is a
time independent quantity. Thus, if ripple coarsening takes place in a given
experiment, the finally observed wavelength is expected to depart from the
value given by \eqref{l1d}.

\paragraph{Velocity of transverse ripple motion} \label{LVTRM}

A third pattern feature that we can extract analytically within linear
approximation is the velocity for transverse ripple motion. This is the velocity at which, say, a local minimum of the linear ripple structure travels across the substrate, corresponding to the {\em phase} velocity of a wave packet.\cite{mattheij:2005} Note that the imaginary part of the dispersion relation only depends on the $x$ component of the wave-vector, so that (linear) ripple motion takes place only in the $x$ direction.
In order to compute its velocity we simply have to take the ratio between the imaginary part of the linear dispersion relation and the wave-vector, evaluating at the maximum of the real part of $\omega_k^{}$. Thus,
\begin{equation}
V^\ell =  \frac{{\rm Im}(\omega_k^{})}{k_x} \Big|_{k_x^{\ell}} 
 = \phi \alpha_0 \alpha_{1x} + \frac{4 \pi^2\alpha_0}{({l}^\ell)^2} \left(-\phi
 \alpha_{3x} + \Delta_x \alpha_{1x}\right) . \label{v1d_b}
\end{equation}
In the case of one-field models, an analogous expression is obtained, except
for the new term proportional to $\Delta_x$, that appears here due to
the coupling between erosion and transport. Note the importance of an
analogous term (that is proportional to the ion beam flux and whose final sign
is opposed to that of the combined first and second summands in \eqref{v1d_b},
see $e.g.$ Appendix \ref{app.A}) in order to correctly account for the
experimental direction of ripple motion, as stressed in
[\onlinecite{alkemade:2006}]. In this reference, thermal spikes were invoked in
order to justify such an extra contribution. In contrast, our present two-field
formulation allows to obtain a similar correction [$e.g.$ in the analogous zero redeposition limit we get $V^\ell = \alpha_0 \alpha_{1x} - (4 \pi^2\alpha_0/({l}^\ell)^2) (\alpha_{3x} + R_{eq} \gamma_{2x} \alpha_{1x})$], without the need for mechanisms that differ from, say, linear collision cascades combined with surface transport. Nevertheless, as with the ripple wavelength, nonlinear effects can in general influence the observed velocity of lateral ripple motion, as seen in Sec.\ \ref{Sub_Non-lineal_regime}.

\section{Nonlinear analysis and effective interface equation} \label{nonlinear}

During the development of the morphological instability, a time is reached
after which nonlinear terms can no longer be neglected and a nonlinear analysis
is needed. Note that the band of unstable Fourier modes extends from
$k^*=\sqrt{2}k^{\ell}_{x,y}$ down to $k=0$, its size being controlled by the square
root of the small parameter $\epsilon$, as seen in Eq.\ \eqref{kl}. Moreover,
the fastest growing mode $k^{\ell}$ is also proportional to $\epsilon^{1/2}$. Thus,
$\epsilon^{-1/2}$ provides us with a characteristic length scale associated
with the linear instability and makes it natural to define slow spatial
variables that are of order unity at the scale of the linear ripple wavelength,
namely, $X=\epsilon^{1/2}x$ and $Y=\epsilon^{1/2}y$. Moreover, it is also
possible to obtain a estimation of the time scales associated with the
translation (the imaginary part of $\omega_{\mathbf{k}}^+$) and growth (the
real part of $\omega_{\mathbf{k}}^+ $) of the linear instability. Thus, by
substituting the value of $k^{\ell}$ in \eqref{ImW} and \eqref{ReW}, the imaginary
part scales as $\epsilon^{3/2}$ and the real part as $\epsilon^2$. Hence,
analogously to the slow spatial variables, we can define two slow time
variables, $T_1=\epsilon^{3/2}t$ and $T_2=\epsilon^2 t$, associated with
in-plane translation and vertical growth, respectively. These natural
variables will allow us to perform a multiple-scale analysis in order to obtain
a closed equation for $h$ using the fact that, near the instability threshold
(namely, for small $\epsilon$ values), $R$ tends to its stationary value much
faster than $h$. This will be seen to allow for an adiabatical (perturbative)
elimination of $R$ from the dynamics.

We will use a frame of reference comoving with the planar solution [Eqs.\
\eqref{Rp} and \eqref{hp}] in order to investigate how the solution evolves around it. We write
\begin{eqnarray}
h&=&h^p+\widetilde{h} \\
R&=&R^p+\widetilde{R}.
\end{eqnarray}
The strategy consists in expanding $\widetilde{h}$ and $\widetilde{R}$ in
powers of $\epsilon^{1/2}$, substituting these expressions into Eqs.\
\eqref{eq.R} and \eqref{eq.h}, and solving to increasingly higher
orders in $\epsilon$. Before doing that we will write Eqs.\ \eqref{eq.R}
and \eqref{eq.h} in terms of the slow, $X$, $Y$, $T_1$, and $T_2$
variables by means of the chain rule
\begin{align}
\partial_x&= \epsilon^{1/2} \partial_X, \label{Dx} \\
\partial_y&=\epsilon^{1/2} \partial_Y, \label{Dy}\\
\partial_t &= \epsilon^{3/2}\partial_{T_1} + \epsilon^{2} \partial_{T_2},
\label{Dt}
\end{align}
to obtain
\begin{align}
    \epsilon^{3/2} \partial_{T_1} \widetilde{R} + \epsilon^{2} \partial_{T_2} \widetilde{R}
&= \bar{\phi} \widetilde{{\Gamma}}_{ex} - \widetilde{{\Gamma}}_{ad}  + \epsilon
D
    \nabla^2 \widetilde{R} \label{eq.Rb},\\
     \epsilon^{3/2} \partial_{T_1} \widetilde{h} + \epsilon^{2} \partial_{T_2} \widetilde{h}
&= - \widetilde{{\Gamma}}_{ex} + \widetilde{{\Gamma}}_{ad} \label{eq.hb},
\end{align}
with
\begin{align}
\widetilde{\Gamma}_{ad}&= \gamma_0 \left[ \widetilde{R} + \epsilon R_{eq}
\nabla \cdot
(\underline{\gamma_2}\nabla \widetilde{h}) \right], \label{Gaepsilon}\\
\widetilde{\Gamma}_{ex}&= \gamma_0 R_{eq} \left\{\epsilon^{3/2} \alpha_{1x}
\partial_X \widetilde{h} \right.  \nonumber\\
&+ \epsilon^2 [\nabla \cdot (\underline{\alpha_2} \nabla \widetilde{h})
+ \nabla \widetilde{h} \cdot (\underline{\alpha_6} \, \nabla \widetilde{h})] \nonumber \\
&+ \epsilon^{5/2} [\partial_X \nabla \cdot (\underline{\alpha_3} \nabla
\widetilde{h}) +
\partial_X \widetilde{h} \, \nabla \cdot (\underline{\alpha_5} \nabla \widetilde{h})] \nonumber \\
&+ \left. \epsilon^3 \nabla \cdot (\underline{\alpha_4} \nabla \nabla^2
\widetilde{h}) \right\}, \label{Geepsilon}
\end{align}
where we have used the value of the temporal derivatives of the planar
solutions, $R^p$ and $h^p$, given by \eqref{eq.Rp} and \eqref{eq.hp}, expressed
all space derivatives in the slow variables, and defined $\underline{\gamma_2}
\equiv \textrm{diag}(\gamma_{2x},\gamma_{2y})$.

Expanding now $\widetilde{R}$ and $\widetilde{h}$ in powers of $\epsilon^{1/2}$
as
\begin{align}
\widetilde{R} &= \sum_{n=0} \epsilon^{n/2} \, R_n , \label{expansionR} \\
\widetilde{h} &= \sum_{n=0} \epsilon^{n/2} \, h_n , \label{expansionh}
\end{align}
we seek to solve for the various orders $R_n$, $h_n$ by substituting the above
expansions into \eqref{eq.Rb} and \eqref{eq.hb}.

Note that from Eq.\ \eqref{eq.Rb} and substituting the value of
$\widetilde{\Gamma}_{ad}$ we obtain
\begin{multline}
\widetilde{R} = - \epsilon R_{eq}  \nabla \cdot (\underline{\gamma_2} \nabla
\widetilde{h}) + \frac{1}{\gamma_0} \Big(\bar{\phi}\,
 \widetilde{\Gamma}_{ex}+ \epsilon D \nabla^2 \widetilde{R} \\
-\epsilon^{3/2} \partial_{T_1} \widetilde{R} - \epsilon^{2} \partial_{T_2}
\widetilde{R} \Big)  , \label{eq.Rnew}
\end{multline}
which, together with the shape of $\widetilde{\Gamma}_{ex}$ given by
\eqref{Geepsilon}, indicates that, for any order $n$, the $R_n$ coefficient
depends on terms of lower orders in the expansion of $\widetilde{R}$ and
$\widetilde{h}$. Thus, the terms obtained in the expansion of $\widetilde{R}$
can be substituted back into \eqref{eq.hb} to get a closed equation for the
evolution of $\widetilde{h}$. While details of this procedure are given in
Appendix \ref{app.C}, the resulting equation reads, in the original time and
space variables,
\begin{multline}
\partial_t h = \partial_t h^p + \gamma_x \partial_x h  \\
 + \sum_{i=x,y} \left[\nu_i \, \partial^2_i h + \lambda^{(1)}_{i} \,
 (\partial_i h)^2 + \Omega_ i \partial_i^2 \partial_x h
 + \xi_i\, (\partial_x h) (\partial^2_i h)\right]  \\
 -\sum_{i,j=x,y} \left[ {\mathcal K}_{ij} \partial^2_i \partial^2_j h +
 \lambda^{(2)}_{ij}\, \partial^2_{i} (\partial_j h)^2 \right], \label{eq.ero}
\end{multline}
where we have neglected height derivatives that are of sixth or higher orders,
we have undone the transformation to the frame comoving with the planar
solution, and parameters are related to those of the original two-field model
\eqref{eq.R}-\eqref{eq.h} as
\begin{align}
  \gamma_x&=-\alpha_0 \phi \alpha_{1x}, \nonumber \\
  \nu_x&= -\alpha_0 \phi \alpha_{2x} + \frac{\alpha_0^2}{\gamma_0} \bar{\phi} \phi  \alpha_{1x}^2,
  \quad \nonumber \nu_y = - \alpha_0 \phi \alpha_{2y},\\
  \lambda_{i}^{(1)}&=- \alpha_0 \phi \alpha_{6i}, \quad \xi_i =-\alpha_0 \phi \alpha_{5i}, \nonumber\\
  \Omega_i& = \alpha_0 \left[-\phi \alpha_{3i}+ \left(\frac{\bar{\phi}D}{\gamma_0}
  - \phi R_{eq} \gamma_{2i}\right) \alpha_{1x}\right], \nonumber \\
  {\mathcal K}_{ij}&= D R_{eq} \gamma_{2i} + \alpha_0 \left[\phi\alpha_{4ij}-
  \left( \frac{ \bar{\phi} D}{\gamma_0} - \phi R_{eq} \gamma_{2i} \right)
  \alpha_{2j}\right] , \nonumber \\
  \lambda^{(2)}_{ij}&=-\alpha_0 \left( \frac{ \bar{\phi} D}{\gamma_0} -
  \phi R_{eq} \gamma_{2i} \right) \alpha_{6j}. \label{relacion_eq.ero}
\end{align}
Note that in \eqref{relacion_eq.ero} we have restored the expression of
$\epsilon$ in terms of physical parameters. As mentioned in Sec.\ \ref{Sub_planar},
after a time of order $\gamma_0^{-1}$ the the profile erodes uniformly with
velocity $v_0=-\partial_t h^p=\phi \alpha_0$.

We have obtained a closed evolution equation for $h$ from which $R$ has been
eliminated, and whose behavior is equivalent to that predicted by the full
two-field model near the instability threshold. Note that, in particular, the
linear dispersion relation for \eqref{eq.ero} coincides, within our long
wavelength approximation, with that of the original model as given by
\eqref{ImW} and \eqref{ReW}. Moreover, as in previous one-field descriptions,
in Eq.\ \eqref{eq.ero} there is not reflection symmetry in the $x$ direction
due to the oblique ion incidence. This symmetry is restored if the bombardment
is perpendicular to the substrate, or else if the target is rotated
simultaneously with irradiation, as described in [\onlinecite{munoz-garcia:2007b}]. Actually, Eq.\
\eqref{eq.ero} generalizes the anisotropic interface equation \eqref{TNLeqh}
that is obtained by one-field theories\cite{makeev:2002} by the appearance of
additional nonlinear terms (with coefficients $\lambda^{(2)}_{ij}$). These,
together with the modified dependence of parameters on physical constants, are
the main effects of having explicitly described the dynamics of the diffusive
field $R$ onto the evolution of the profile, and will be seen below to be
instrumental in order to provide an improved description of nanopatterning by
IBS.

\section{1D model} \label{1-D_model}

Eq.\ \eqref{eq.ero} is a highly non-linear and anisotropic system whose full
analysis is rather complex. Before analyzing it in detail, and in order to
understand more directly the physical content of its various terms and
parameter dependences on physical constants, we are going to study first a 1D
counterpart of the erosion model studied in previous sections. We will thus
consider that the $x$ axis is the only relevant direction to describe the
topography of the system. This simplification is very frequently done in models
for sand ripples formation,\cite{terzidis:1998,valance:1999,csahok:2000} in
which translation invariance is assumed in the direction perpendicular
to the wind. Note that such an approximation still respects the physically
essential lack of reflection symmetry in the $x$ axis.
Thus, by repeating the approach of the previous section in the case that there is no variation of the fields in the $y$ direction, we obtain the following one-dimensional equation
\begin{multline}\label{eq.ero1d}
\partial_t h  =  -v_0+\gamma_x \partial_x h + \nu_x \partial^2_x h +
\lambda^{(1)}_{x} (\partial_x h)^2 + \Omega_x \partial_x^3 h \\
  +  \xi_x\,
(\partial_x h) (\partial^2_x h)- {\mathcal K}_{xx} \partial^4_x h -
\lambda^{(2)}_{xx}\, \partial^2_{x} (\partial_x h)^2,
\end{multline}
where by an abuse of language we will employ similar symbols
for parameters to those of the previous Section, and the relation of these with
the coefficients of the coupled model are
\begin{align}
  &v_0=\alpha_0 \phi; \, \gamma_x=-\alpha_0 \phi \alpha_{1x}; \,
  \nu_x= -\alpha_0 \phi \alpha_{2x} + \frac{\alpha_0^2}{\gamma_0} \bar{\phi} \phi  \alpha_{1x}^2;
 \nonumber \\
    &\Omega_x  = \alpha_0 \left[-\phi \alpha_{3x}+ \left(\frac{\bar{\phi}D}{\gamma_0}
  - \phi R_{eq} \gamma_{2x}\right) \alpha_{1x}\right];
  \xi_x =-\alpha_0 \phi \alpha_{5x}; \nonumber \\
  &{\mathcal K}_{xx}= D R_{eq} \gamma_{2x} + \alpha_0 \left[\phi\alpha_{4xx}- \left( \frac{ \bar{\phi} D}{\gamma_0}
  - \phi R_{eq} \gamma_{2x} \right) \alpha_{2x}\right] ; \nonumber \\
  & \lambda_{x}^{(1)}=- \alpha_0 \phi \alpha_{6x}; \, \lambda^{(2)}_{xx}=-\alpha_0 \left( \frac{ \bar{\phi} D}{\gamma_0} - \phi R_{eq}
  \gamma_{2x} \right) \alpha_{6x}. \label{relacion_eq.ero1d}
\end{align}
Eq.\ \eqref{eq.ero1d} provides the generalization of the 1D counterpart of Eq.\
\eqref{TNLeqh}, through appearance of the additional $\lambda^{(2)}_{xx}$ term.
Actually, restricting ourselves to even terms in $x$ derivatives (that is, for $\gamma_x=\Omega_x=\xi_x=0$), Eq.\ \eqref{eq.ero1d} becomes the mixed
Kuramoto-Sivashinsky equation (see [\onlinecite{munoz-garcia:2006b}] and
references therein) that generalizes the Kuramoto-Sivashinsky (KS) equation.\cite{kuramoto:1976,sivashinsky:1977}
In general note that the coefficients \eqref{relacion_eq.ero1d} directly reproduce those associated with the $x$ direction among the larger set of parameters in \eqref{relacion_eq.ero}. Although one dimensional, Eq.\ \eqref{eq.ero1d} is still a highly nonlinear equation with behaviors that may range from in-plane traveling periodic (ordered) structures to chaotic (disordered) cell dynamics, as occurs with its $\lambda^{(2)}_{xx}=0$ limit.\cite{bar:1995,castro:2007}

\subsection{Physical interpretation of parameters}

Before attempting to understand the interplay among the various terms in Eq.\
\eqref{eq.ero1d}, it is worth giving consideration to each one of them
individually. To this end, it is instructive to start by studying the two
possible limiting cases for parameter $\phi$.

\subsubsection{Complete redeposition ($\phi=0$)} \label{sec.fi0}

Equation \eqref{eq.ero1d} becomes strongly simplified when the erosive
mechanism limits itself to transferring material from the immobile bulk to the
mobile diffusive current, without sputtering proper, akin to the role of IBS
for ion beam assisted deposition.\cite{rauschenbach:2002} In this case, the
only non-zero coefficients in \eqref{relacion_eq.ero1d} are
\begin{align}
  \Omega_x & = \frac{\alpha_0 D  \alpha_{1x}}{\gamma_0} ,\quad \lambda^{(2)}_{xx} =-\frac{\alpha_0 D \alpha_{6x}}{\gamma_0} , \nonumber \\
  {\mathcal K}_{xx}&= D R_{eq} \gamma_{2x} - \frac{\alpha_0 D
  \alpha_{2x}}{\gamma_0} , \label{relacion_1d.fi0}
 \end{align}
the interface equation reading merely
\begin{equation}\label{eq.1dfi0}
\partial_t h = \Omega_x \partial_x^3 h - {\mathcal K}_{xx} \partial^4_x h -
\lambda^{(2)}_{xx}\, \partial^2_{x} (\partial_x h)^2.
\end{equation}
This equation has the conserved form expected from the fact that excavation is
here limited to matter redistribution. Actually, in the absence of the third
order derivative term, Eq.\ \eqref{eq.1dfi0} in known as the conserved KPZ
equation,\cite{lai:1991,barabasi,cuerno:2004} relevant to conserved surface
growth dynamics such as in typical Molecular Beam Epitaxy systems. Note
that, although the surface diffusion coefficient $\mathcal{K}_{xx}$ of Eq.\
\eqref{relacion_1d.fi0} includes an erosive contribution that is of a
destabilizing nature as long as excavation is favored at surface minima
($\alpha_{2x}>0$), being proportional to $\alpha_0$ this contribution is
numerically smaller than the stabilizing (thermal) contribution also present in
$\mathcal{K}_{xx}$. The only remaining nonlinearity in \eqref{eq.1dfi0}
reflects (through $\alpha_{6x}$) the non-linear dependence of the excavation
rate with the local surface slope. Moreover, already this term genuinely
couples erosion to transport, being also proportional to $D$.

\subsubsection{Zero redeposition ($\phi=1$)} \label{sec.fi1}

This limit corresponds to the usual assumption in previous one-field
approaches. In this case generically Eq.\ \eqref{eq.ero1d} displays its full
shape, with coefficients
\begin{align}
  v_0&=\alpha_0; \quad \gamma_x=-\alpha_0 \alpha_{1x}; \quad
 \nu_x= -\alpha_0 \alpha_{2x} \nonumber \\
 \Omega_x & = -\alpha_0 \left(\alpha_{3x}+ R_{eq} \gamma_{2x} \alpha_{1x} \right); \quad
  \xi_x =-\alpha_0 \alpha_{5x}; \nonumber \\
  {\mathcal K}_{xx}&= D R_{eq} \gamma_{2x} + \alpha_0 \left(\alpha_{4xx} + R_{eq} \gamma_{2x} \alpha_{2x}
  \right) ; \nonumber \\
  \lambda_{x}^{(1)}&=- \alpha_0 \alpha_{6x}; \quad \lambda^{(2)}_{xx}=\alpha_0
  R_{eq} \gamma_{2x} \alpha_{6x}. \label{relacion_1dfi1}
\end{align}
Among coefficients in \eqref{relacion_1dfi1}, all but three of them
($\Omega_x$, ${\cal K}_{xx}$, and $\lambda^{(2)}_{xx}$) are directly as
predicted by one-field models, see \eqref{relacion_Ge}. As for the three
remaining coefficients, common to all three is that they correspond to
conservative terms in the equation of motion. This allows to understand the
contributions that they include in which transport (through dependence on
$R_{eq}$) couples to an erosive dependence on a height derivative two orders
lower. $E.g.$ $\Omega_{x}$ is associated with a third order height derivative
and indeed features a direct erosive dependence in the 3rd.\ order coefficient
$\alpha_{3x}$. However, it also depends (through $R_{eq}$) on the first order
erosive coefficient $\alpha_{1x}$. Similarly for ${\cal K}_{xx}$ and
$\lambda^{(2)}_{xx}$. The surface diffusion coefficient ${\cal K}_{xx}$ adds to these the
expected contribution $D R_{eq} \gamma_{2x}$ discussed in Sec.\ \ref{sec.Mullins}.
Moreover, note that the ion effective smoothing term with coefficient
$\alpha_{4xx}$, that reflects the dependence of the excavation rate with high
(fourth) order surface derivatives, appears as a direct contribution to the
surface diffusion coefficient. 

About the coefficient of the conserved KPZ term, note that for this $\phi=1$ case its sign is opposite to
that of $\lambda_x^{(1)}$ in \eqref{relacion_1dfi1}. This leads to a
cancellation mode and mathematically invalidates Eq.\ \eqref{eq.ero1d} as a
description of the physical system. Indeed, neglecting the $\xi_x$ nonlinearity
that does not participate in the height saturation of the linear
instability,\cite{csahok:2000} the remaining nonlinear contributions read, in
Fourier space, $-(\lambda^{(1)}_x + k^2 \lambda_{xx}^{(2)}) {\cal F}[(\partial_x h)^2]$, where ${\cal F}$ denotes Fourier transform.
Due to the signs of the coefficients, there is a wave vector in the
unstable band (cancellation mode) for which the parenthesis in this equation
vanishes, rendering the system {\em nonlinearly} unstable.\cite{raible:2001}
This undesirable feature actually also occurs in full 2D one-field models when
generalized to sufficiently high orders.\cite{kim:2004,castro:2005b,kim:2005}

\subsubsection{Partial redeposition ($0<\phi<1$)}

Generically we expect partial redeposition to occur under
usual experimental conditions for IBS nanopatterning. After the previous
Section, we see that not only is redeposition a physical effect to
include but also that it allows to regularize our mathematical description of
the system. Indeed, using the parameter combination defined in \eqref{eq.Deltas},
we see that parameter conditions exist for small but non-zero values $0 < \phi
< 1$, for which $\Delta_x > 0$ so that $\lambda^{(1)}_x$ and
$\lambda^{(2)}_{xx}$ have the same sign and cancellation modes do not occur.
The numerical values of $\phi$ and $\Delta_x$ also affect the remaining
coefficients in \eqref{relacion_eq.ero1d}, but are of a less critical nature. The only contribution that is privative of these partial
redeposition conditions is the second term in the expression for $\nu_x$, that,
being positive, is of a stabilizing nature and opposes the sputtering
instability. A similar term can be found in the formation of macroscopic
ripples under the action of the wind when the number of sand grains is {\em
not} conserved,\cite{misbah:2003} and reflects the geometrical fact that
erosion tends to smooth out inclined surface features. Nevertheless, such a
term being higher order in powers of $\alpha_0$, we expect it to be numerically
small in most practical cases within our IBS context. In general, the $0 < \phi
< 1$ case interpolates between the two extreme cases considered above, in that
the dependence of coefficients \eqref{relacion_eq.ero1d} on physical parameters
combine the features discussed in Secs.\ \ref{sec.fi0} and \ref{sec.fi1}.

\subsection{Effective interface equation vs full two-field model}
\label{2Dvs1D}

In order to check the analytical approximations made in the derivation of the
effective interface equation and compare its predictions on the dynamics to
those of the full original two-field model, we have performed a numerical
integration of the 1D coupled set of Eqs.\ \eqref{eq.R}-\eqref{eq.h}, and of the related single Eq.\ \eqref{eq.ero1d}, using an Euler scheme for the time integration, and the improved spatial discretization introduced by Lam
and Shin \cite{lam:1998} for the nonlinear terms. We have used periodic
boundary conditions, lattice constant $\delta x = 1$ and time step
$\delta t = 0.01$, checking that results do not differ significantly
for smaller space and time steps. The standard system size of our simulation
has been $L=2048$. With the aim of comparing the evolution of the profile for
the two-field and the effective equations, the same random initial height values were chosen, uniformly distributed between $- 0.05$ and $0.05$, and the
corresponding parameters were related following \eqref{relacion_eq.ero1d}.

\begin{figure}[!htmbp]
\begin{center}
\includegraphics[width=0.48\textwidth]{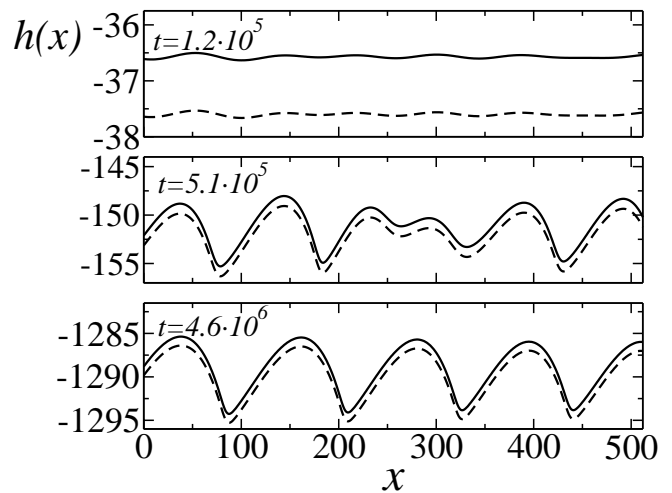}
\caption{Height profiles $h$ (dashed line) and thickness of the mobile material over $h$ (solid line) 
at different times given by the 1D two-field coupled model with $\bar{\phi}=0.99$, $\alpha_0=0.03$, $\alpha_1=-1$ $\alpha_2=30$, $\alpha_3=\alpha_4=1$, $\alpha_5=-1$, $\alpha_6=-3$, $R_{eq}=\gamma_0=\gamma_2=1$, and $D=10$. All units are arbitrary. \label{fig1}}
\end{center}
\end{figure}

We show in Fig.\ \ref{fig1} the evolution described by the 1D two-field model of the height profile $h$ and the thickness of the
mobile material above $h$ for certain values of the parameters.
Since $\epsilon=3\cdot10^{-3}$ is small, we see that $R$ is indeed only slightly
altered from its equilibrium value ($R_{eq}=1$).
Note how the morphological instability leads to formation of a periodic ripple
pattern that, as expected, is not symmetric in the $x$ direction. The
thickness of the mobile surface layer correlates with the topography all along the dynamics, being smaller at steeper ripple sides.

\begin{figure}[!htmbp]
\begin{center}
\includegraphics[width= 0.48\textwidth]{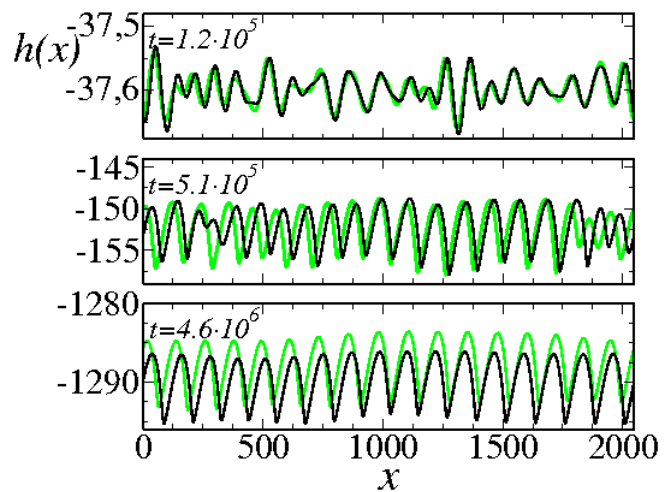}
\caption{(color online) Height profiles at different times given by the 1D two-field model (black line) for parameters as in
Fig.\ \ref{fig1} and by the effective equation [Eq.\ \eqref{eq.ero1d}] (green
line) with parameters as given by relation \eqref{relacion_eq.ero1d},
namely, $v_0=-3\cdot10^{-4}$, $\gamma_x=3\cdot10^{-4}$, $\nu_x=-9\cdot10^{-3}$,
$\lambda_{x}^{(1)}= 9\cdot10^{-4}$, $\Omega_x=-0.297$ , $\xi_x=3\cdot10^{-4}$,
$\mathcal{K}_{xx}=1.0993$, and $\lambda^{(2)}_{xx}=0.8901$.
All units are arbitrary. \label{fig2}}
\end{center}
\end{figure}

In Fig.\ \ref{fig2} we compare the evolution of the profile for the 1D two field
coupled model, with that described by the effective height equation, Eq.\ \eqref{eq.ero1d}, where the
coefficients of both systems are related by \eqref{relacion_eq.ero1d}.
We can see how, starting from the same flat random initial distribution for
both systems, a periodic surface structure appears with a wavelength of about the maximum of the linear dispersion relation, and the amplitude of height variations increases. For the examples considered in Fig.\ \ref{fig2} the wavelength of the linear instability is given by \eqref{l1d}, yielding ${l}^\ell = 98$. At these short times, when the slopes are not too large so that nonlinear terms are not yet relevant, both profiles match quite accurately. Far from the linear
instability threshold, the profiles become less similar. Since the space and time scales separation and the power expansion performed to obtain the effective equation are only valid for small values of $\epsilon$, it is expected that, the smaller $\epsilon$ is, the more similar the profiles become. However, if we reduce this parameter, the simulations are more time consuming since the characteristic space and time scales for pattern dynamics are inversely proportional to powers of $\epsilon$, as noted in Sec.\ \ref{nonlinear}. In
any case, for the values of $\epsilon$ considered in our simulations, the effective equation captures the main features of the original two-field
model, even in terms of the behavior of observables such as the global surface rms width or roughness $W(t)$ or the ripple wavelength $l(t)$ (defined as the
mean lateral distance between two consecutive local minima), as seen in Figs.\ \ref{fig3} and \ref{fig4}, respectively.


\subsection{Nonlinear dynamics for the effective equation}

Indeed, at later stages, nonlinearities determine the evolution of the surface morphology. For the reasons mentioned above, we will explore this regime through the effective interface equation. Specifically, nonlinear effects induce coarsening of structures wherein the cells (ripples) grow in width and height, their number decreasing in both systems. For both cases coarsening is such that smaller cells are ``eaten" by larger neighbors until reaching constant amplitude and wavelength values, while lateral moundlike order is still preserved for intermediate distances (more than ten times the lateral size of the cells). This behavior is very similar to that reported in Ref.\ [\onlinecite{munoz-garcia:2006b}] for the mixed KS equation equation that corresponds to the $\gamma_x =\Omega_x=\xi_x=0$ limit of \eqref{eq.ero1d}; see also paper II.

\begin{figure}[!htmbp]
\begin{center}
\includegraphics[width=0.48\textwidth]{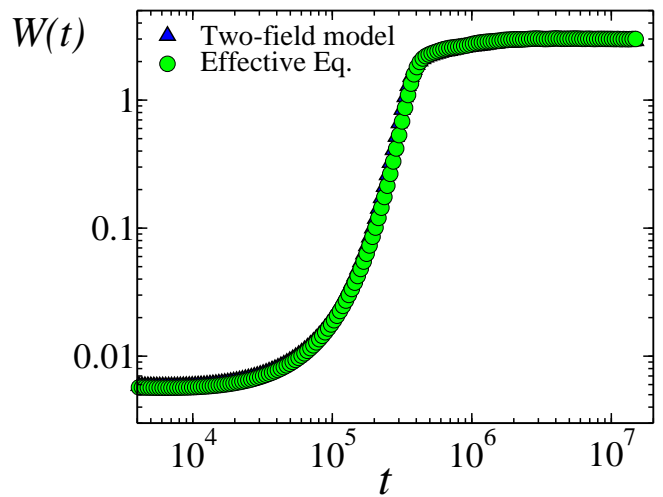}
\caption{(color online) Temporal evolution of the global roughness, $W(t)$ given by the two-field model (blue triangles) and by the effective interface equation (green circles) for the same coefficients as in Fig.\ \ref{fig2}. Error bars are smaller than the symbol sizes. All units are arbitrary. \label{fig3}}
\end{center}
\end{figure}

\begin{figure}[!htmbp]
\begin{center}
\includegraphics[width= 0.48\textwidth]{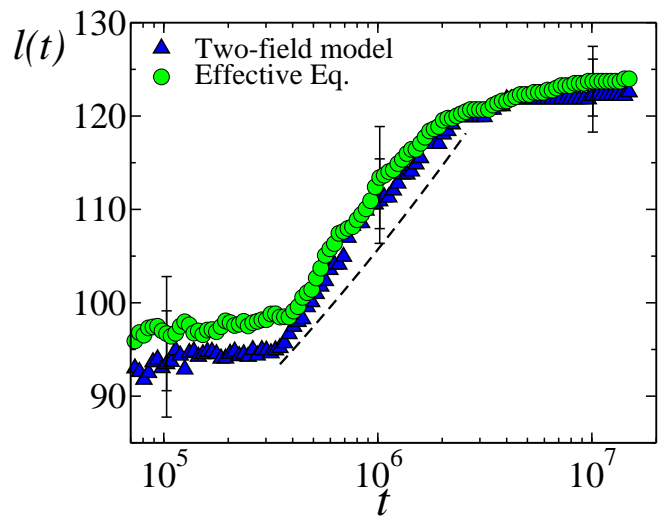}
\caption{(color online) Temporal evolution of the lateral pattern wavelength, ${l}(t)$, given by the two-field model (blue triangles) and by the effective interface equation (green circles) for the same coefficients as in Fig.\ \ref{fig2}. A few representative error bars are given that represent statistical dispersion. The dashed line corresponds to ${l}(t) \sim t^{0.12}$. All units are arbitrary. \label{fig4}}
\end{center}
\end{figure}

In Figs.\ \ref{fig3} and \ref{fig4} we show the time evolution of the surface  roughness $W(t)$ and ripple wavelength ${l}(t)$, averaged over 18 random initial
conditions. After a stage in which the amplitude of the linear instability and,
therefore $W$, grow exponentially, a coarsening process begins (roughly at $t\simeq 3 \cdot 10^5$) for the ripple wavelength. Around $t=2\cdot 10^6$ this process stops and the wavelength and amplitude of the pattern reach stationary values. Specifically, the lateral pattern wavelength grows from its initial value corresponding to the linear instability ${l}^\ell=98$ until a saturation value, close to ${l}=121$. At intermediate times this coarsening behavior can be described by an effective power law ${l} \sim t^{0.12}$, as suggested in Fig.\ \ref{fig4}. In the presence of coarsening, the dependence of the asymptotic values of the ripple amplitude and wavelength with system parameters differs from those of the linear instability regime. If one assumes \cite{csahok:2000} that the odd-derivative terms in Eq.\ \eqref{eq.ero1d} do not contribute to such a coarsening process, approximate values can be obtained through comparison with coarsening dynamics in the conserved KS equation.\cite{munoz-garcia:2006b} Such estimates are more accurate in the normal incidence case (paper II),\cite{munoz-garcia:2007b} to which we refer the interested reader.
Additional important features of these systems, which are not present in the
equation studied in Ref.\ [\onlinecite{munoz-garcia:2006b}], are the asymmetry of the profile and the lateral movement of the pattern. As we have checked in our simulations, the asymmetry on the pattern depends only on the (advective) terms corresponding to the coefficients $\Omega_x$ and $\xi_x$ of the effective
equation. For negative values of $\Omega_x$ and/or positive values of $\xi_x$
the cell structure tends to be leaning to the right. This can be observed in
Figs.\ \ref{fig1} and \ref{fig2}, where the right slopes of the cells are
clearly larger than the left slopes. If $\Omega_x$ is positive and/or $\xi_x$ is negative, the pattern is leaning to the left. If both terms have the same sign, the orientation of the structure depends on their relative magnitude. 

\begin{figure}[!htmbp]
\begin{center}
\includegraphics[width=0.48\textwidth]{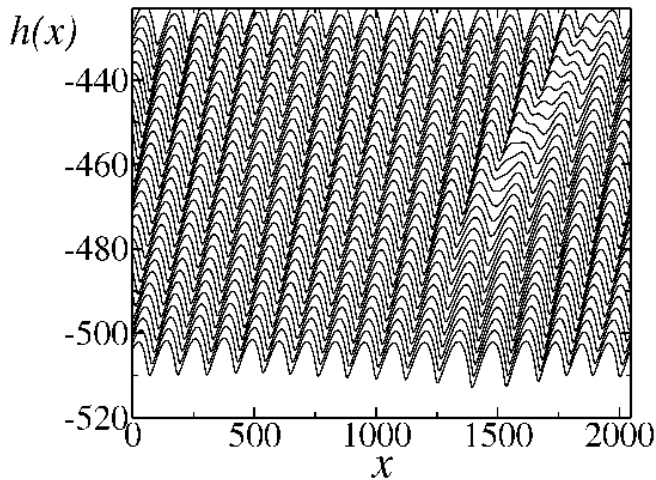}
\caption{Height profiles between $t= 1.5\cdot 10^{6}$ (top) and
$t=1.8\cdot10^6$ (bottom) evaluated at equally spaced intervals of $10^4$ time units, for the effective equation \eqref{eq.ero1d}, and parameters as in Fig.\ \ref{fig2}. All units are arbitrary. \label{fig5}}
\end{center}
\end{figure}

Considering lateral ripple motion, note first that the linear prediction for the velocity, Eq.\ \eqref{v1d_b} has the form $V^\ell = -\gamma_x + 4 \pi^2 \Omega_x({l}^\ell)^{-2}$. The contribution due to $\gamma_x$ is an uniform translation (along a direction on the $x$ axis that is opposite to the sign of $\gamma_x$) that can actually be cancelled out by an appropriate choice of reference frame. Thus, the only remaining terms which influence in-plane displacement of the pattern are again $\Omega_x$ (linear) and $\xi_x$ (non-linear).  For the values we have considered for the remaining parameters, a positive sign of $\Omega_x$ and/or $\xi_x$ induces ripple motion towards positive $x$, while negative values of these parameters lead to lateral ripple motion in the opposite direction. In Fig.\ \ref{fig5} we observe the lateral movement of the pattern as described by Eq.\ \eqref{eq.ero1d}, for parameters as in Fig.\ \ref{fig2}. Simultaneously with erosion and mean height evolution towards larger negative values, the pattern is moving towards the left. Here, the movement is dominated by $\gamma_x$ and $\Omega_x$, which induce motion towards the negative $x$ values.

The results reported in this section allow us to conclude that both the effective interface equation and the two field model, whose coefficients are related through (\ref{relacion_eq.ero1d}), capture common features observed in experiments such as the coarsening process of the pattern wavelength, the short range lateral order and the non uniform lateral displacement of the structure. On the other hand, due to the fact that the scales associated with the experimental linear instability are very large (of the order of $\epsilon^{-1/2}$), one needs very large simulations in order to compare with experiments. These are available to the effective equation, in which parameters can be rescaled with the aim of accelerating the simulations. For these reasons, in going to the physical 2D case in the next section, we will limit our study to the 2D effective height equation \eqref{eq.ero}. We will consider some illustrative examples of the ensuing surface dynamics that allow us to understand the richness of the behaviors that can be described by such a complex non-linear system.

\section{Full 2D effective interface equation}\label{2-D_equation}

Equation \eqref{eq.ero} generalizes the one-dimensional equation \eqref{eq.ero1d} to the case of fully anisotropic two-dimensional targets, in a way that is consistent with reflection symmetry in the $y$ direction, as expected from the ion incidence geometry. As mentioned earlier, Eq.\ \eqref{eq.ero} generalizes the one-field equation \eqref{TNLeqh}, through appearance here of the (anisotropic) conserved-KPZ type terms $\partial_i^2(\partial_j h)^2$. In turn, Eq.\ \eqref{TNLeqh} already provided an anisotropic generalization (through the presence of odd derivatives in the $x$ coordinate) of the two-dimensional KS equation.\cite{cuerno:1995,rost:1995} To the best of our knowledge, Eq.\ \eqref{eq.ero} is new and adds to the relatively small number \cite{bar:1995} of (local) evolution equations for fully anisotropic two-dimensional pattern forming systems, that are derived from constitutive laws. In the context of hydrodynamic models of ripple formation on aeolian sand dunes, an isotropic 2D equations, when available, are limited to conservative dynamics,\cite{yizhaq:2004} while in the cases of thin film surfaces 
nonlinearities that arise are of a different type.\cite{sato:1999,golovin:1999,levandovsky:2006}

Although the parameter space of Eq.\ \eqref{eq.ero} is much larger than that of its one-dimensional counterpart \eqref{eq.ero1d}, the physical interpretation of the various terms and coefficients is completely analogous, corresponding to a natural generalization of those appearing in the latter. Given that the main linear features of the two-dimensional equation were already discussed (and compared with typical experimental data) in Sec.\ \ref{linear_inst}, we next consider numerical simulations of Eq.\ \eqref{eq.ero} that show the main morphological features of its full dynamics, that will be later compared with experimental results. Some peculiarities on the cancellation modes that may arise in Eq.\ \eqref{eq.ero} are considered analytically in a specific subsection.


\subsection{2D dynamics: numerical results} \label{2-D equation}

Far from a complete analysis of Eq\ \eqref{eq.ero}, we will limit ourselves in this section to a qualitative study of its main properties and how it successfully reproduces some experimental features which are not included in previous continuum descriptions. 

Thus, we have performed a numerical integration of Eq.\ \eqref{eq.ero} using 
an scheme that generalizes that employed in the one-dimensional case, namely, an
Euler updating rule with $\delta t=10^{-3}$ for the time evolution, and the finite difference prescription of [\onlinecite{lam:1998}] for the nonlinear terms, with $\delta x=1$. The standard size of our
simulations was $L\times L = 256 \times 256 $ with periodic boundary and random initial conditions. We consider a reference plane comoving with the eroded surface with a constant velocity $-v_0$, thus the effective equation that we integrate is \eqref{eq.ero} for $v_0=0$.

\begin{figure}[!htmb]
\begin{center}
\begin{minipage}[c]{0.49\linewidth}
\begin{center}
\includegraphics[width=\linewidth,angle=90]{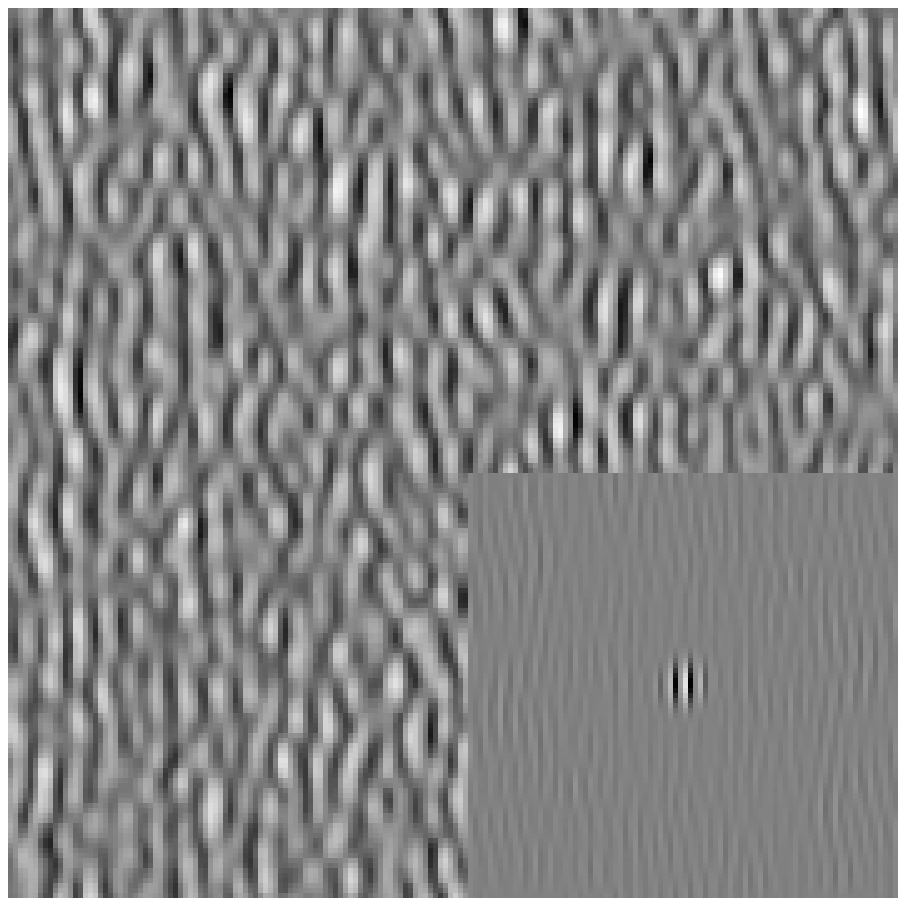}
\end{center}
\end{minipage}\hspace*{ 0.015\linewidth}
\begin{minipage}[c]{0.49\linewidth}
 \begin{center}
\includegraphics[width=\linewidth]{Compressed_FigurasEPS/fig6d.eps}
\end{center}
\end{minipage}
\vspace{0.03\linewidth}\\
{\large (a)}
\vspace{0.06\linewidth}\\
\begin{minipage}[c]{0.49\linewidth}
\begin{center}
\includegraphics[width=\linewidth,angle=90]{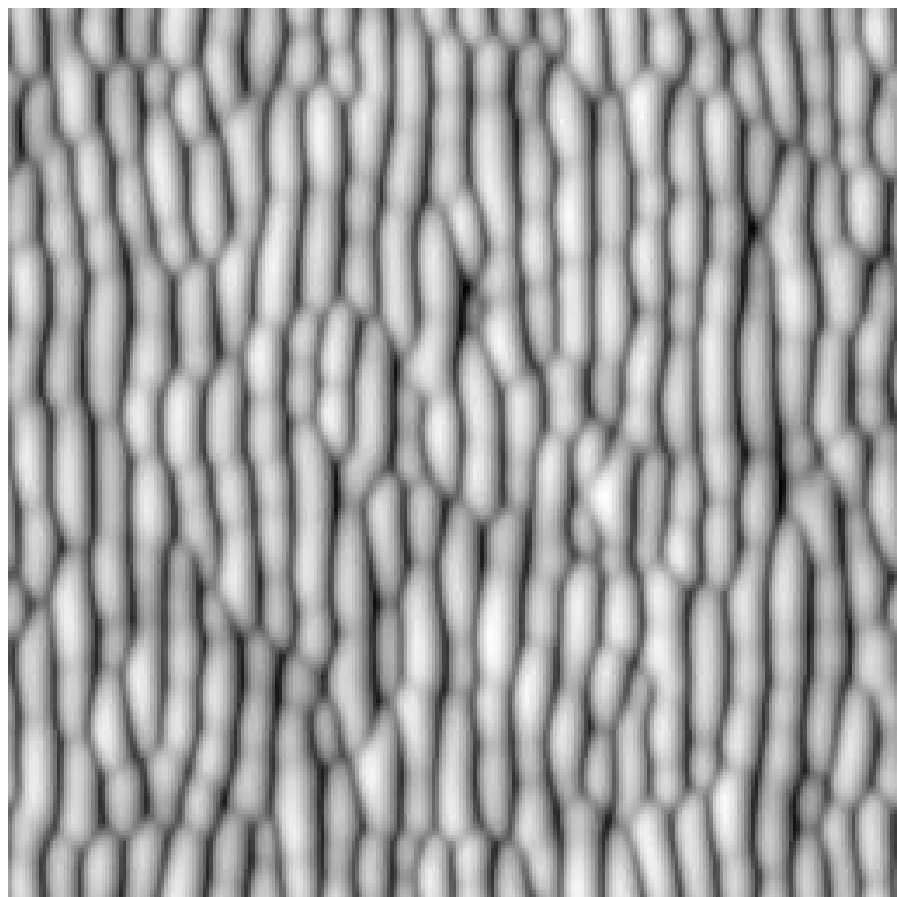}
\end{center}
\end{minipage}\hspace*{ 0.015\linewidth}
\begin{minipage}[c]{0.49\linewidth}
 \begin{center}
\includegraphics[width=\linewidth]{Compressed_FigurasEPS/fig6e.eps}
\end{center}
\end{minipage}
\vspace{0.03\linewidth}\\
{\large (b)}
\vspace{0.06\linewidth}\\
\begin{minipage}[c]{0.49\linewidth}
\begin{center}
\includegraphics[width=\linewidth,angle=90]{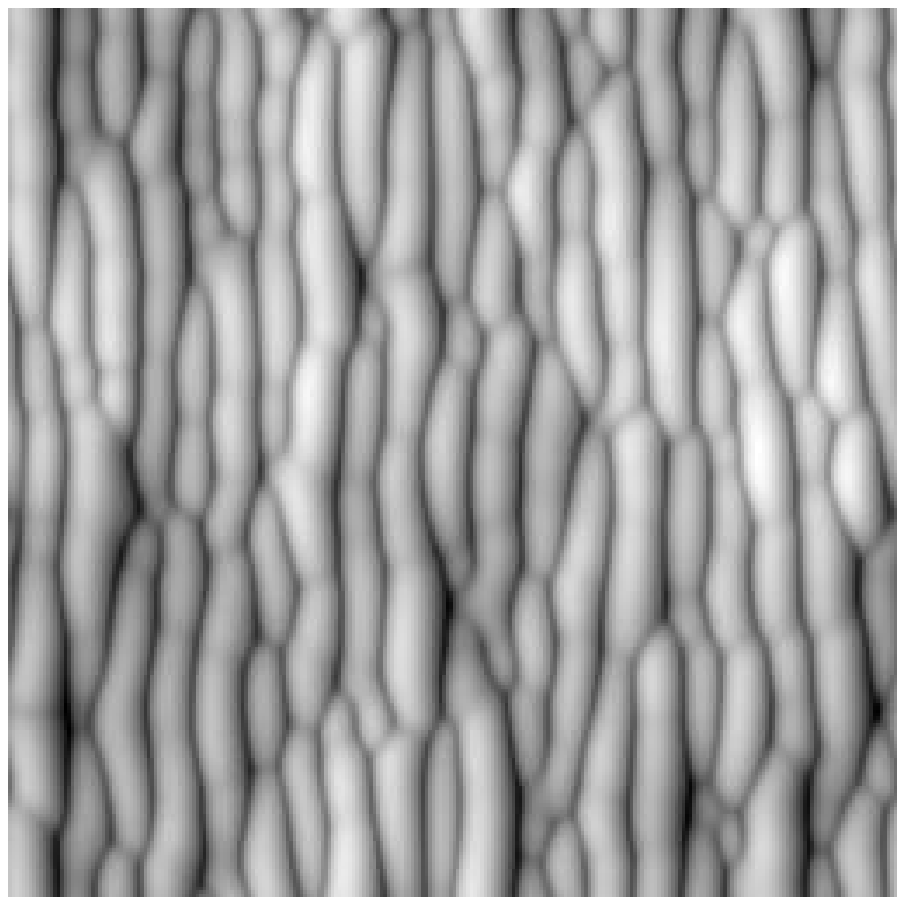}
\end{center}
\end{minipage}\hspace*{ 0.015\linewidth}
\begin{minipage}[c]{0.49\linewidth}
 \begin{center}
\includegraphics[width=\linewidth]{Compressed_FigurasEPS/fig6f.eps}
\end{center}
\end{minipage}
\vspace{0.03\linewidth}\\
{\large (c)}
\vspace{0.03\linewidth}\\
\caption{Time evolution of relatively disordered ripples with mild wavelength coarsening (see also supplementary video).\cite{EPAPS1} Snapshots at increasing times: (a) $t=10$; (b) $t=106$; (c) $t=953$ for Eq.\ \eqref{eq.ero} with $v_0=0$, $\gamma_x=-0.1$, $\nu_x=-1$, $\nu_y=-0.1$, $\Omega_x=1$, $\Omega_y=0.5$,
$\xi_i=0.1$, $\lambda_x^{(1)}=1$, $\lambda_y^{(1)}=5$, $\lambda^{(2)}_{i,j}=5$,
and ${\cal K}_{i,j}=1$. Top views (left column) and corresponding transverse
cuts at $y=L/2$ (right column). Inset in (a) is its corresponding height autocorrelation. All units are arbitrary. \label{oblicua_r5}}
\end{center}
\end{figure}

\begin{figure}[!htmb]
\begin{center}
\begin{minipage}[c]{0.49\linewidth}
\begin{center}
\includegraphics[width=\linewidth,angle=90]{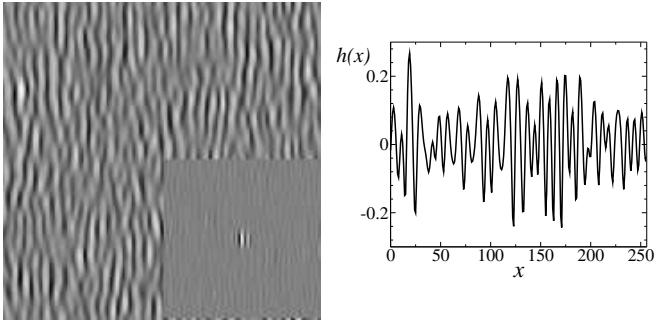}
\end{center}
\end{minipage}\hspace*{ 0.015\linewidth}
\begin{minipage}[c]{0.49\linewidth}
 \begin{center}
\includegraphics[width=\linewidth]{Compressed_FigurasEPS/fig7d.eps}
\end{center}
\end{minipage}
\vspace{0.03\linewidth}\\
{\large (a)}
\vspace{0.06\linewidth}\\
\begin{minipage}[c]{0.49\linewidth}
\begin{center}
\includegraphics[width=\linewidth,angle=90]{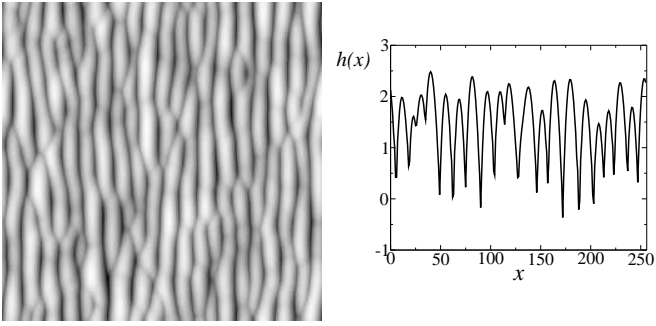}
\end{center}
\end{minipage}\hspace*{ 0.015\linewidth}
\begin{minipage}[c]{0.49\linewidth}
 \begin{center}
\includegraphics[width=\linewidth]{Compressed_FigurasEPS/fig7e.eps}
\end{center}
\end{minipage}
\vspace{0.03\linewidth}\\
{\large (b)}
\vspace{0.06\linewidth}\\
\begin{minipage}[c]{0.49\linewidth}
\begin{center}
\includegraphics[width=\linewidth,angle=90]{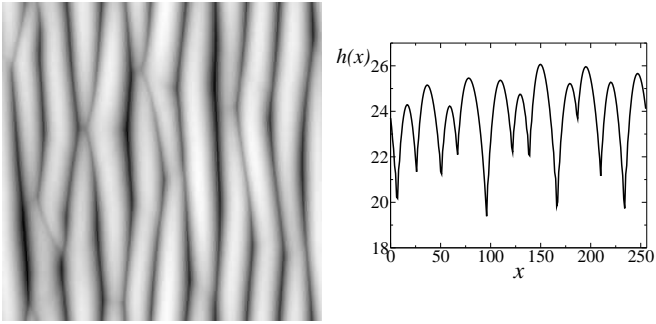}
\end{center}
\end{minipage}\hspace*{ 0.015\linewidth}
\begin{minipage}[c]{0.49\linewidth}
 \begin{center}
\includegraphics[width=\linewidth]{Compressed_FigurasEPS/fig7f.eps}
\end{center}
\end{minipage}
\vspace{0.03\linewidth}\\
{\large (c)}
\vspace{0.03\linewidth}\\
\caption{Time evolution of relatively ordered ripples with sizeable wavelength coarsening (see also supplementary video).\cite{EPAPS2} Snapshots at increasing times: (a) $t=10$; (b) $t=106$; (c) $t=953$ for Eq.\ \eqref{eq.ero} with the same parameters
as in Fig.\ \ref{oblicua_r5}, except for $\lambda_x^{(1)}=0.1$. Top views (left
column) and corresponding transverse cuts at $y=L/2$ (right column). Inset in (a) is its corresponding height autocorrelation. All units are arbitrary. \label{oblicua_r50}}
\end{center}
\end{figure}

\begin{figure}[!htmb]
\begin{center}
\begin{minipage}[c]{0.49\linewidth}
\begin{center}
\includegraphics[width=\linewidth,angle=90]{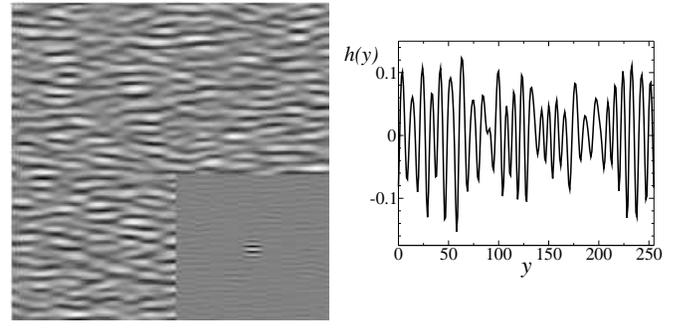}
\end{center}
\end{minipage}\hspace*{ 0.015\linewidth}
\begin{minipage}[c]{0.49\linewidth}
 \begin{center}
\includegraphics[width=\linewidth]{Compressed_FigurasEPS/fig8d.eps}
\end{center}
\end{minipage}
\vspace{0.03\linewidth}\\
{\large (a)}
\vspace{0.06\linewidth}\\
\begin{minipage}[c]{0.49\linewidth}
\begin{center}
\includegraphics[width=\linewidth,angle=90]{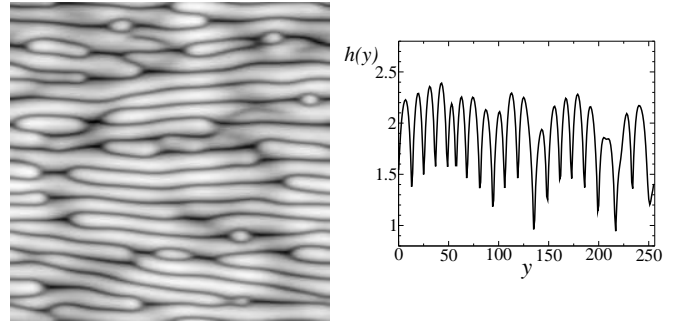}
\end{center}
\end{minipage}\hspace*{ 0.015\linewidth}
\begin{minipage}[c]{0.49\linewidth}
 \begin{center}
\includegraphics[width=\linewidth]{Compressed_FigurasEPS/fig8e.eps}
\end{center}
\end{minipage}
\vspace{0.03\linewidth}\\
{\large (b)}
\vspace{0.06\linewidth}\\
\begin{minipage}[c]{0.49\linewidth}
\begin{center}
\includegraphics[width=\linewidth,angle=90]{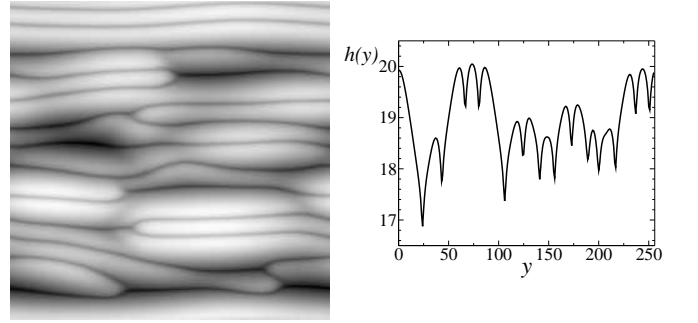}
\end{center}
\end{minipage}\hspace*{ 0.015\linewidth}
\begin{minipage}[c]{0.49\linewidth}
 \begin{center}
\includegraphics[width=\linewidth]{Compressed_FigurasEPS/fig8f.eps}
\end{center}
\end{minipage}
\vspace{0.03\linewidth}\\
{\large (c)}
\vspace{0.03\linewidth}\\
\caption{Time evolution of relatively disordered ripples with sizeable wavelength coarsening (see also supplementary video).\cite{EPAPS3} Ripple orientation as for typical large incidence angle conditions. Snapshots at increasing times: (a) $t=10$; (b) $t=106$; (c) $t=953$ for Eq.\ \eqref{eq.ero} with $v_0=0$,
$\gamma_x=0.1$, $\nu_x=1$, $\nu_y=-0.95$, $\Omega_i=-0.5$, $\xi_i=0.1$,
$\lambda_x^{(1)}=0.1$, $\lambda_y^{(1)}=1.0$, $\lambda^{(2)}_{i,x}=0.5$,
$\lambda^{(2)}_{i,y}=5.0$, and ${\cal K}_{i,j}=1$. Top views (left column) and
corresponding transverse cuts at $x=L/2$ (right column). Inset in (a) is its corresponding height autocorrelation. All units are arbitrary. \label{oblicua_nu+1}}
\end{center}
\end{figure}

The evolution of the height as described by Eq.\ \eqref{eq.ero} is depicted in
Figs.\ \ref{oblicua_r5}-\ref{oblicua_nu+1} for different values of the
coefficients, with the $x$-axis oriented along the horizontal direction (see also supplementary videos).\cite{EPAPS1,EPAPS2,EPAPS3} In each figure three snapshots (top views and lateral cuts) are provided for a given parameter condition, with time increasing from panel (a) to panel (c).
In all these examples, and resembling experimental morphologies,\cite{munoz-garcia:2007} both the 
amplitude and the wavelength of the ripples grow with time, while the pattern disorders in heights for long lateral distances. The detailed shapes of the topographies, however, are quite different depending on the values of the
parameters. We can obtain longitudinally disordered ripples which are frequently
interrupted along the direction of the crests, as in Fig.\ \ref{oblicua_r5}, or else ordered straight and wide ripples occur for different parameter conditions as in Fig.\ \ref{oblicua_r50}. An even more disordered pattern is
depicted in Fig.\ \ref{oblicua_nu+1}, where the ripples group into domains of
about three cells whose crests run along the $x$ axis, as expected from the parameter values (note $\nu_x > 0$ in this example). 

Similarly to the one-dimensional case, before slopes are large enough to make non-linear terms non-negligible, the evolution of the morphology is governed by linear terms. This will allow us to separate the dynamics into two different regimes, linear and nonlinear, according to the type of terms that control the evolution.

\subsubsection{Linear regime}\label{Sub_Lineal_regime}

As noted in Sec.\ \ref{nonlinear}, the linear dispersion relation of Eq.\
\eqref{eq.ero} coincides with that of the original model described in
subsection \ref{Linear_stability_analysis}. Thus, for isotropic thermal surface
diffusion, the ripple crests are oriented along the $y$ ($x$) axis if $\nu_x$
($\nu_y) $ is more negative than $\nu_y$ ($\nu_x$), thus reproducing the ripple
orientation as predicted by the BH theory. Numerical integration within linear
regime indeed retrieves the dependence of the ripple orientation as a function of the values of $\nu_x$ and $\nu_y$ as shown in Figs.\ \ref{oblicua_r5}, \ref{oblicua_r50}, and \ref{oblicua_nu+1}. Furthermore, we have also checked in our simulations that the lateral wavelength of the pattern is given by the relation between the surface tension and diffusion terms.  One way to do that is to measure the distance from the origin to the first maximum in the height autocorrelation function which is represented in the inset of Figs. \ \ref{oblicua_r5}(a), \ref{oblicua_r50}(a), and \ref{oblicua_nu+1}(a). Since ${\mathcal K}_{ij}=1$ is considered for all these examples, we have ${l}^\ell=2\pi/k^{\ell}_i=2\pi(-2/\nu_i)^{1/2}$.

\begin{figure}[!htmbp]
\begin{center}
\includegraphics[width=0.48\textwidth]{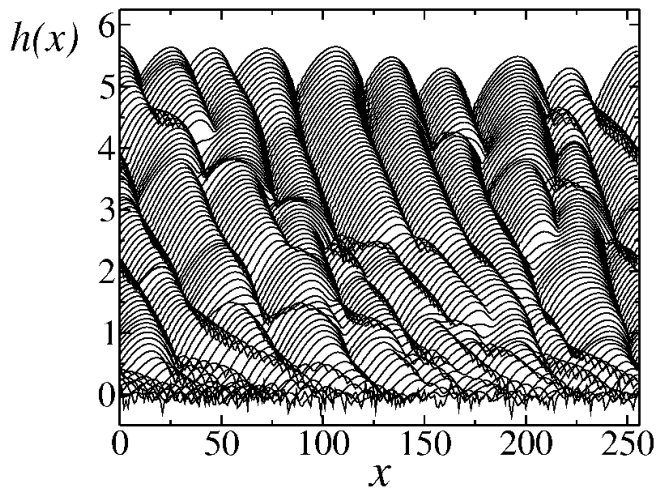}
\caption{In-plane non-uniform ripple motion as seen from the evolution of
transverse cuts of the surface at $y=L/2$ for equally spaced times between $t=0$ and $t=1500$. Results from the numerical integration of Eq.\
\eqref{eq.ero} with $v_0=0$, $\nu_x=-1$, $\nu_y=-0.1$, $\gamma_x=\xi_i=0$,
$\Omega_i=-2$, $\lambda_x^{(1)}=1$, $\lambda_y^{(1)}=5$,
$\lambda^{(2)}_{i,x}=50$, $\lambda^{(2)}_{i,y}=5.0$, and ${\cal K}_{i,j}=1$.
All units are arbitrary. \label{perfiles_Xi0_g0}}
\end{center}
\end{figure}

While even linear derivatives in Eq.\ \eqref{eq.ero} are responsible for
amplification or attenuation of the ripple amplitude, they do not induce lateral motion of pattern. Conversely, odd derivatives breaking the $x \rightarrow -x$ symmetries indeed induce in-plane lateral ripple motion. We have checked in our simulations that, as expected, the term corresponding to the coefficient $\gamma_x$ does not alter the shape of the morphology but merely produces a uniform movement along the $x$ axis. As in the one-dimensional case, the direction of this movement is opposite to the sign of $\gamma_x$. On the other hand,
again as in the 1D case, the $\Omega_i$ terms are responsible for both lateral
movement of the structure and shape asymmetry.  These effects can be observed in Fig.\ \ref{perfiles_Xi0_g0} 
where we show the time evolution (as seen from a comoving reference frame) of transverse cuts of the surface for a given parameter condition. We have checked that, indeed, transforming back to a rest reference frame, the ripple 
velocity coincides, for times within linear regime, with that predicted by Eq.\ \eqref{v1d_b}. Already visual inspection of Fig.\ \ref{perfiles_Xi0_g0} suggests deviations from a uniform velocity for transverse ripple motion. This is a signature of nonlinear effects [specifically, due to ripple coarsening manifested by a non constant ripple wavelength ${l}(t)]$, that are considered next.


\subsubsection{Non-linear regime}\label{Sub_Non-lineal_regime}

\begin{figure}[!htmbp]
\begin{center}
\includegraphics[width=0.48\textwidth]{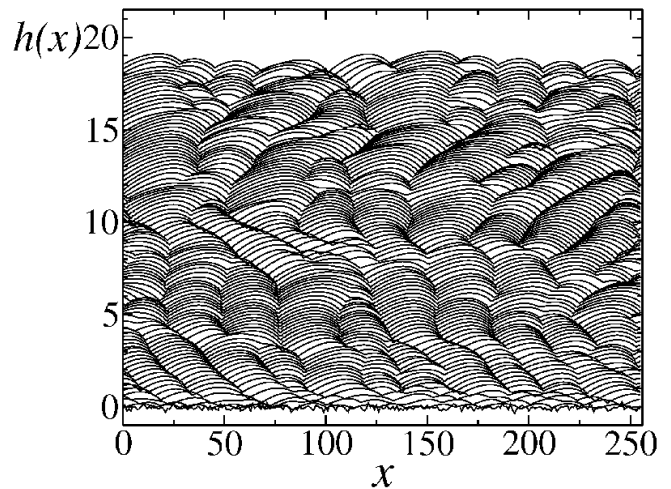}
\caption{Change in the direction of in-plane ripple motion as seen from the evolution of transverse cuts of the surface at $y=L/2$ for equally spaced times between $t=0$ and $t=5000$. Results from the numerical integration of Eq.\
\eqref{eq.ero} for parameters as in Fig.\ \ref{perfiles_Xi0_g0} except
for $\xi_i=4.5$. All units are arbitrary. \label{perfiles_Xi_4.5}}
\end{center}
\end{figure}

For long enough times, non-linear terms have to be considered in order to understand the evolution of the morphology. Those containing even derivatives are reflection symmetric in $x$ and, therefore, are not responsible for lateral movement or any asymmetries of the pattern. On the other hand, we have checked that the terms corresponding to the coefficients $\xi_i$ indeed induce lateral motion of the pattern and asymmetry in the $x$ axis. For the parameters considered in our simulations, positive values of $\xi_i$ induce a non-uniform lateral motion of the pattern towards positive $x$ values. Since the contributions of the $\xi_i$ nonlinearities to the evolution of $h$ increase in the non-linear regime, these can even induce
a change in the direction of the pattern movement as observed in Fig.\
\ref{perfiles_Xi_4.5}, where we plot the time evolution of a transverse cut of
the surface. In this figure $\Omega_i =-2 < 0$, thus, as noted in the previous subsection, this induces a movement of the pattern towards negative $x$. These terms dominate during the linear regime but, as a result of the increase of the values of lower order surface derivatives, the $\xi_i =4.5 > 0$ terms take over and change the direction of lateral ripple motion towards positive $x$ values. This example underscores the complex ripples dynamics induced by nonlinear effects, that should be taken into account in the discussion of the potential limitations of the current BH picture to quantitatively describe ripple motion.\cite{chan:2007,alkemade:2006} 

\begin{figure}[!htmbp]
\begin{center}
\includegraphics[width=0.48\textwidth]{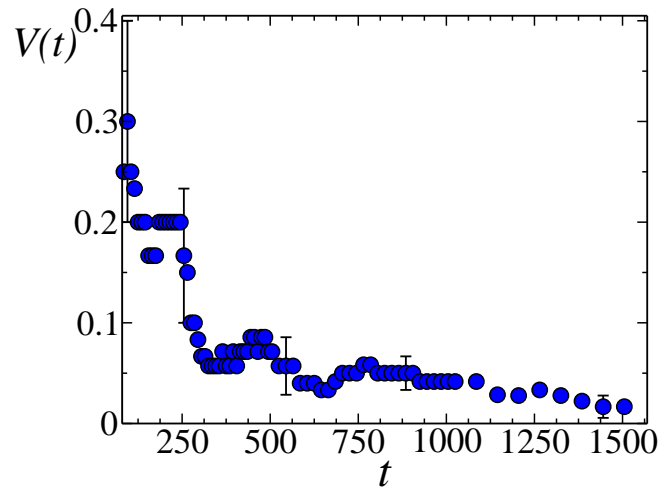}
\caption{(color online) Temporal evolution of the lateral ripple velocity, $V(t)$, given by the numerical integration of Eq.\ \eqref{eq.ero} in the non-linear regime for the same coefficients as in Fig.\ \ref{perfiles_Xi0_g0}. A few representative error bars are given. All units are arbitrary. \label{velocidad}}
\end{center}
\end{figure}

A simpler type of non-uniform ripple motion that has been reported experimentally corresponds to movement in a fixed direction, but with a non-uniform velocity, see $e.g.$ [\onlinecite{habenicht:2002,alkemade:2006}]. As mentioned above, this behavior correlates with the occurrence of wavelength coarsening (see below), and Eq.\ \eqref{eq.ero} is the first two-dimensional continuum equation to describe it within the IBS context. As an example, in Fig.\ \ref{velocidad} we show the (non-uniform) ripple velocity $V(t)$ in the non-linear regime as a function of time for the same simulations as shown in Fig.\ \ref{perfiles_Xi0_g0}. Here the velocity is computed for a single surface minimum once the pattern is completely formed. At longer times the ripple velocity seems to reach a negligible value compatible with arrest of ripple motion. This might be related with a similar interruption of ripple coarsening that is illustrated below.

\begin{figure}[!htmbp]
\begin{center}
\includegraphics[width=0.48\textwidth]{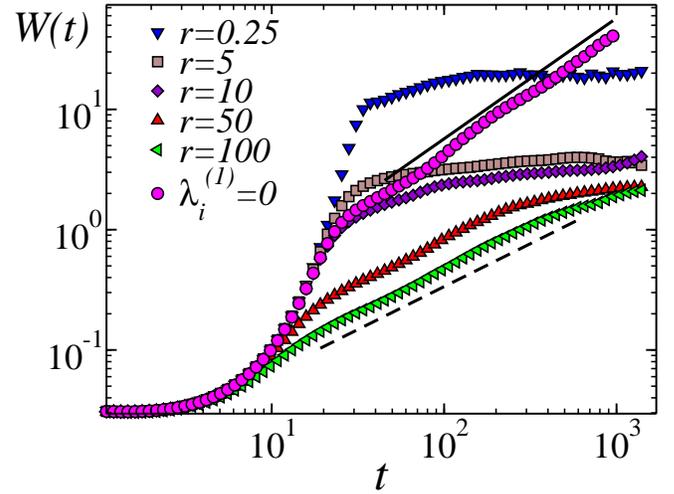}
\caption{(color online) Temporal evolution of the global roughness, $W(t)$, given by the the
effective equation Eq.\ \eqref{eq.ero} with $v_0=0$, $\nu_x=-1$, $\nu_y=-0.1$,
$\gamma_x=\xi_i=\Omega_i=0$, ${\cal K}_{i,j}=1$, $\lambda_i^{(1)}=0.1$, and
$\lambda^{(2)}_{i,j}=0.1r$, for different values of $r$. The solid and dashed lines show
the fit to power laws for $\lambda_i^{(1)}=0$ and $r=100$ where $W \sim t^{1.00}$ and $W \sim t^{0.71}$, respectively. All units are arbitrary. \label{Rugosidad}}
\end{center}
\end{figure}

\begin{figure}[!htmbp]
\begin{center}
\includegraphics[width= 0.48\textwidth]{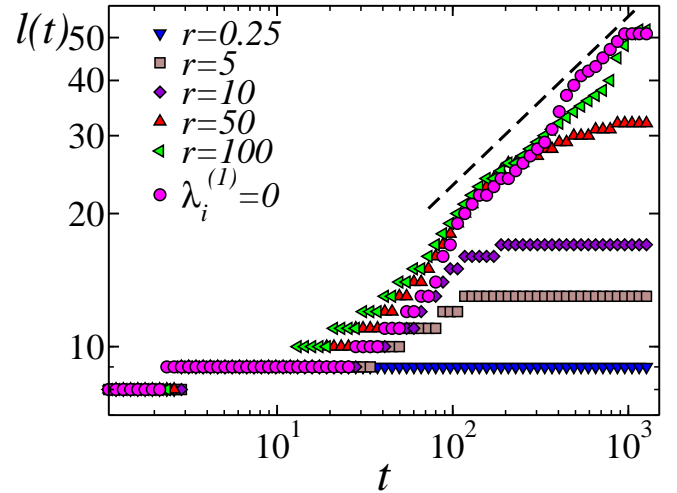}
\caption{(color online) Temporal evolution of the lateral wavelength of the pattern, ${l}(t)$, given by the effective equation Eq.\ \eqref{eq.ero} for the same coefficients as Fig.\ \ref{Rugosidad}. The dashed line shows
the fit to a power law for $r=100$ where ${l} \sim t^{0.38}$. The fit in the same region for $\lambda_i^{(1)}=0$ yields ${l} \sim t^{0.48}$ (not shown). All units are arbitrary. \label{oblicua_l}}
\end{center}
\end{figure}

Non-linear terms containing derivatives that are reflection-symmetric in $x$ are responsible for the eventual saturation of the ripple amplitude, and for the quality and range of in-plane order of the ripple pattern. As checked in
our simulations and described for the 1D effective equations studied in Sec.\
\ref{2Dvs1D} and [\onlinecite{munoz-garcia:2006b}], the larger the value of the
ratio of $\lambda^{(2)}_{i,j}$ to $\lambda^{(1)}_{i}$ terms is, the longer is the coarsening process, and the more ordered the morphology becomes for the same total time ($i.e.$, ion dose). This is shown in Figs.\ \ref{Rugosidad} and \ref{oblicua_l}, in which the time evolution of the global
surface roughness and of the lateral wavelength of the pattern are depicted
for different values of this ratio, denoted as $r$. In general, the roughness increases exponentially (linear instability regime), after which the nonlinearities are able to stabilize the system and induce slower growth for the roughness, $W(t)$ finally reaching a time independent value. For very small $r$ ratios, this stationary state seems to be reached earlier, and the intermediate slow (power-law) growth regime of the roughness is shorter. For larger values of $r$, this intermediate regime has a wider duration, and can be more accurately described by a power law with the form $W(t) \sim t^{\beta}$ for some effective value of the growth exponent $\beta$. Note that, in the $r\to\infty$ limit (equivalently, $\lambda^{(1)}_{i}$ = 0), Eq.\ \eqref{eq.ero} does {\em not} seem to have a stationary state, similarly to the conserved KS equation.\cite{raible:2000b,munoz-garcia:2006b} Note, the growth exponent for this case is \cite{raible:2000b} $\beta_{\rm cKS}=1$. The gradual chan
 ge of the duration of this intermediate power-law regime with physical parameters (that enter the value of the ratio $r$) and the different values for the effective growth exponent that can be obtained when trying to fit a power-law to such type of data, may account for the spread in the related growth exponents experimentally reported in the context of ripple formation (see references in [\onlinecite{chan:2007,munoz-garcia:2007}]). 

Regarding the quality and range of order in the ripple pattern at intermediate and long times, Fig.\ \ref{Rugosidad} already shows that the morphology is more disordered (the roughness is larger) for smaller values of $r$. Moreover, for these cases, as seen in Fig.\ \ref{oblicua_l}, the stationary value of the pattern wavelength is smaller, and is achieved earlier. A qualitatively similar behavior has been experimentally found in IBS of silicon targets under normal incidence conditions.\cite{gago:2006b} Note, the standard one-field continuum equation \eqref{TNLeqh} corresponds to the $r=0$ limit, for which there is no coarsening and the system is roughest (the roughness being larger almost by an order of magnitude, as seen in Fig.\ \ref{Rugosidad}). Hence, such an equation was not able to account for the observed ripple coarsening and improved ordering, in marked contrast with the present Eq.\ \eqref{eq.ero}. As in the case of the roughness, for the opposite cKS-type limit $r
 \to\infty$ ($\lambda^{(1)}_{i}=0$), the ripple wavelength does not reach a stationary value. Rather, both the amplitude and ${l}(t)$ increase indefinitely, similarly to the cKS case for which ${l}(t)_{\rm cKS} \sim t^{0.5}$, until a single ripple (with a parabolic cross-section) remains in a finite system.\cite{raible:2000b}


Results obtained for the one dimensional anisotropic equation 
\eqref{eq.ero1d} and for the 1D and 2D isotropic counterparts \cite{munoz-garcia:2006b,munoz-garcia:2007b} lead us to expect disorder to dominate the morphological features at the largest length and time scales in the system, as long as cancellation modes do not arise (namely, as long as $\lambda^{(1)}_x$ and $\lambda^{(1)}_y$ have the same signs, see next section). Thus, we expect scale invariant morphologies and rough surfaces for much larger distances than the pattern wavelength. The statistics of the surface fluctuations at these scales are expected to be characterized by the critical exponents of some of the universality classes of kinetic roughening.\cite{barabasi} However, the case of the isotropic KS equation not being even completely understood,\cite{boghosian:1999} we can only conjecture, by analogy with the 1D case, that the asymptotic scaling of Eq.\ \eqref{eq.ero1d} is in the 2D KPZ universality class.

\subsubsection{Cancellation Modes} \label{cancelation_modes}

Eq.\ \eqref{eq.ero} can display cancellation modes (CM), analogously to its own 1D counterpart, Eq.\ \eqref{eq.ero1d}, and to the anisotropic KS (aKS) equation.\cite{rost:1995,park:1999} Recall that CM in Eq.\ \eqref{eq.ero1d} arise due to cancellation between the nonconserved $(\partial_x h)^2$ and the conserved $\partial^2_x(\partial_x h)^2$ KPZ nonlinearities, and lead to (possibly) finite time blow-up of the solutions to the differential equation. We will refer to these as mixed KS (mKS) CM. In marked contrast, CM in the aKS system appear only when the coefficients of the two nonlinear terms $\lambda^{(1)}_x$ and $\lambda^{(1)}_y$ have different signs, and lead to a long time ripple pattern that is oriented along an oblique direction in the $xy$ plane,\cite{rost:1995,park:1999} the system apparently supporting such type of solution for long times. We will denote these as aKS CM. 

Given the large parameter space of Eq.\ \eqref{eq.ero}, the two types of CM mentioned can arise, and we consider separately the conditions for appearance of each of them. Notice, it suffices to consider the nonlinearities that are reflection symmetric in $x$, as they are the only ones involved in the evolution (and putative blow up) of the ripple amplitude.

\paragraph{mKS-type CM.}

The nonconserved and conserved KPZ nonlinearities in Eq.\ \eqref{eq.ero} read
explicitly
\begin{eqnarray}
{\cal N}[h] & \equiv & \lambda^{(1)}_x (\partial_x h)^2 + \lambda^{(1)}_y (\partial_y h)^2 - \lambda^{(2)}_{xx} \partial^2_x(\partial_x h)^2
\label{kpzs} \\
 & & -
\lambda^{(2)}_{xy} \partial^2_x(\partial_y h)^2 -
\lambda^{(2)}_{yx} \partial^2_y(\partial_x h)^2 -
\lambda^{(2)}_{yy} \partial^2_y(\partial_y h)^2 , \nonumber
\end{eqnarray}
whose Fourier transform reads
\begin{equation}
{\cal F}({\cal N}[h])=Q_x {\cal F}[(\partial_x h)^2] + Q_x {\cal F}[(\partial_y h)^2] ,
\label{Fkpzs}
\end{equation}
where we have defined
\begin{eqnarray}
Q_x & = & \lambda^{(1)}_x+\lambda^{(2)}_{xx} k_x^2 + \lambda^{(2)}_{yx} k_y^2 , \label{Qx} \\
Q_y & = & \lambda^{(1)}_x+\lambda^{(2)}_{xy} k_x^2 + \lambda^{(2)}_{yy} k_y^2 .
\label{Qy}
\end{eqnarray}
Now, using \eqref{relacion_eq.ero}, we get
\begin{equation}
Q_i = \lambda^{(1)}_i \left(1+\frac{\Delta}{\phi} k^2 \right) \label{Qs2} ,
\end{equation}
where we have assumed isotropy in the surface tension coefficients as done in Appendix \ref{app.B}, $\gamma_{2x}=\gamma_{2y}=\gamma_2$, and introduced $ \Delta=\bar{\phi}D/\gamma_0 -\phi R_{eq} \gamma_{2}$. As a function of system parameters, there are two possibilities:
\begin{itemize}
\item If $\Delta \geq 0$, then $Q_x, Q_y \neq 0$, so that there are no cancellations among nonconserved and conserved KPZ terms along any direction. This is the 2D generalization of the analogous 1D condition discussed in Section \ref{sec.fi1}.
\item If $\Delta < 0$, then cancellation occurs simultaneously in the $x$ and $y$ directions, for all Fourier modes on the circle $|k_{mKS}|=(\phi/|\Delta|)^{1/2}$, and we expect the solutions of Eq.\ \eqref{eq.ero} to diverge for long times. However, as long as we are close to the instability threshold, the putative CM [being $k_{mKS} \sim O(1)$] are outside the band of linearly unstable modes, so that no divergence occurs and Eq.\ \eqref{eq.ero} still provides a mathematically well-defined model.
\end{itemize}

\paragraph{aKS-type CM.}

\begin{figure}[!htmb]
\begin{center}
\begin{minipage}[c]{0.49\linewidth}
\begin{center}
\includegraphics[width=\linewidth,angle=90]{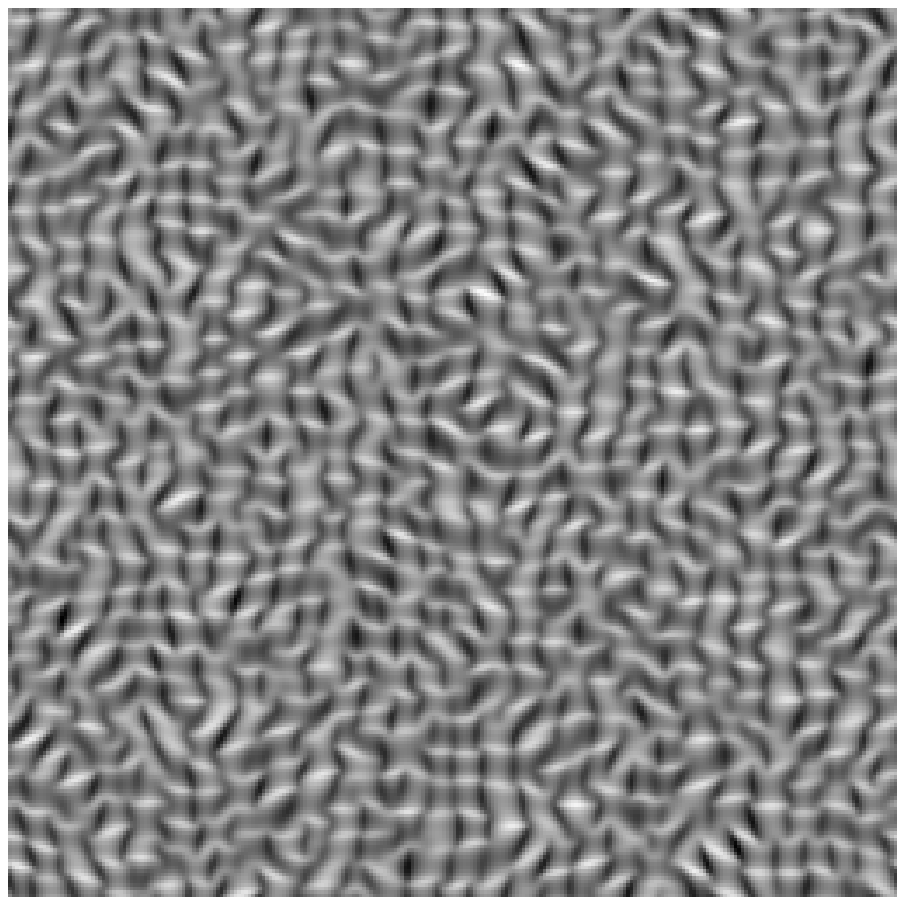}
\vspace{0.02\linewidth} {\large $t=25$}
\end{center}
\end{minipage}\hspace*{ 0.015\linewidth}
\begin{minipage}[c]{0.49\linewidth}
 \begin{center}
\includegraphics[width=\linewidth]{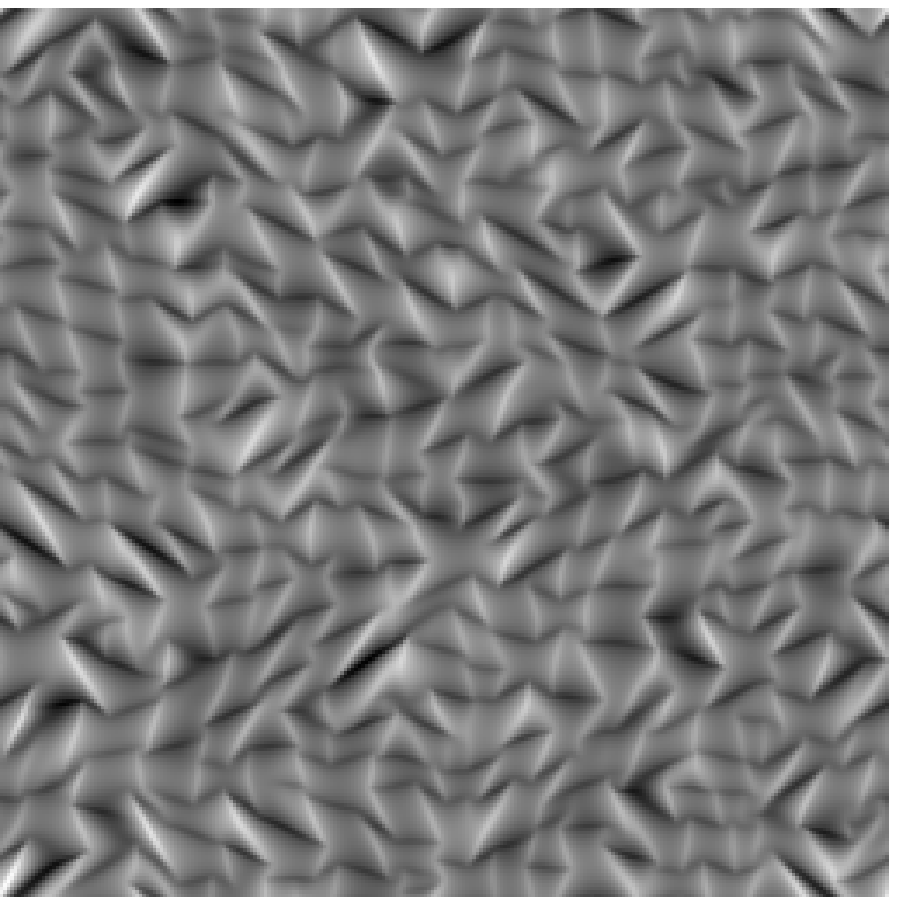}
\vspace{0.02\linewidth} {\large $t=50$}
\end{center}
\end{minipage}
\vspace{0.03\linewidth}\\
\begin{minipage}[c]{0.49\linewidth}
\begin{center}
\includegraphics[width=\linewidth,angle=90]{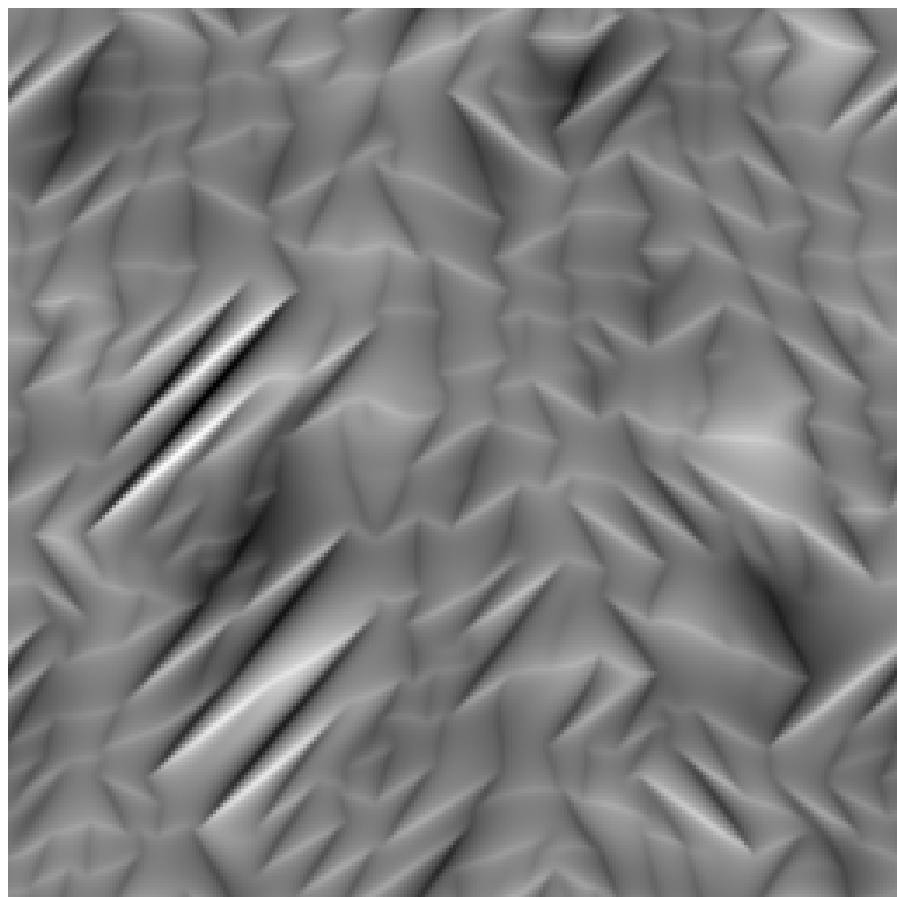}
\vspace{0.02\linewidth} {\large $t=185$}
\end{center}
\end{minipage}\hspace*{ 0.015\linewidth}
\begin{minipage}[c]{0.49\linewidth}
 \begin{center}
\caption{Long time development of oblique ripple patterns at $45^{\circ}$ to the $x$ axis due to cancellation modes (see text). Top views of morphologies obtained by numerical integration of Eq.\
\eqref{eq.ero} with $v_0=0$, $\nu_x=-1$, $\nu_y=1$,
$\gamma_x=\xi_i=\Omega_i=0$, ${\cal K}_{i,j}=1$, $\lambda_x^{(1)}=0.1$,
$\lambda_y^{(1)}=-0.1$, $\lambda^{(2)}_{i,x}=0.5$, and
$\lambda^{(2)}_{i,y}=-0.5$ at increasing times. All units are arbitrary. 
\label{Fig_modos}}
\end{center}
\end{minipage}
\end{center}
\end{figure}

Even in the most favorable case ($\Delta \geq 0$) considered in the previous discussion, there is still the possibility that cancellation takes place, not between nonlinearities of different order (mKS type) but, rather, for specific directions on the $xy$ plane, as in the aKS type. In order to assess such a possibility, we make the Ansatz \cite{rost:1995} that solutions are of the form $h(x,y,t) = f(x-uy,t)$, and see how this reflects into the KPZ nonlinearities \eqref{kpzs}. Thus,
\begin{equation}
{\cal N}[h] = (\lambda^{(1)}_x + u^2 \lambda^{(1)}_y) \left\{ (f')^2 -
\frac{\Delta (1+u^2)}{\phi} [(f')^2]'' \right\} ,
\label{aKS}
\end{equation}
where primes denote differentiation of $f$ with respect to its first argument.
As a consequence, exactly as in the aKS case, whenever the coefficients of the nonconserved KPZ nonlinearities, $\lambda^{(1)}_x$ and $\lambda^{(1)}_y$, have different signs (as a result of their dependence on physical parameters), cancellation takes place for a Fourier mode that is oriented at an angle $\tan^{-1} (-\lambda^{(1)}_x/\lambda^{(1)}_y)^{1/2}$ with the $x$ axis.
Actually, for an appropriate choice of the function $f$ additional cancellation may take place irrespective of the signs of the $\lambda^{(1)}_i$ coefficients, but such special cases are not generic. In Fig.\ \ref{Fig_modos} we show an example of the evolution of the morphology in case of cancellation modes oriented at $45^{\circ}$ to the $x$ axis.
It is tempting to interpret the obliquely oriented ripples recently found \cite{ziberi:2008} on Si at 2 keV as some type of cancellation mode of this aKS type.

\subsection{2D dynamics: comparison to experiments} \label{experiments}

Along the discussion on the detailed surface dynamics predicted by Eq.\ \eqref{eq.ero}, we have already pointed out relation to experimental features that are described by this new effective equation. In our discussion, we have assumed as a reference case that dependencies of coefficients $\alpha_{ikj}$ on the physical parameters such as average ion energy and flux, temperature, and
characteristics of the distribution of energy deposition in the target, are as in BH-type approaches for amorphizable targets. Within such an assumption, all dependencies of {\em linear} features on the latter are as in one-field models,\cite{makeev:2002} whose comparison to experiments has been reviewed in detail elsewhere.\cite{makeev:2002,valbusa:2002,chan:2007,munoz-garcia:2007}
Whenever discrepancies arise, some may be due to deviations of the actual collision cascade statistics from Sigmund's Gaussian formula, and this is a matter of current active research.\cite{feix:2005,davidovitch:2007}

There are other features of our two-field model and of the ensuing Eq.\ \eqref{eq.ero}, that seem more robust to modifications in the values of the parameters entering $\Gamma_{ex}$, provided there is a morphological instability in the ``surface tension'' coefficients. Thus, the formation and fast stabilization of a stationary value for the thickness of the amorphous mobile layer has been assessed $e.g.$ for Si both in Molecular Dynamics \cite{moore:2004} and in experiments, see [\onlinecite{ziberi:2005}] or [\onlinecite{chini:2003}] (for energies of tens of keV). Asymmetry in ripple cross sections has also been assessed both by microscopy (for Si, see [\onlinecite{ziberi:2005}]) or by techniques in reciprocal space ($e.g.$ sapphire \cite{zhou:2007}). Also wavelength coarsening has been profusely documented, there being a large spread in the values of the effective coarsening exponents, see references in [\onlinecite{munoz-garcia:2007}] and more recently $e.g.$ [\onlinecite{katharria:2007}] for SiC. As for in-plane ripple motion, there is a smaller number of studies, although detailed studies (typically employing focused ion beams) are indeed available \cite{habenicht:2002} for Si and for glass,\cite{alkemade:2006} the phenomenon having been reported also in atomistic simulations of amorphous carbon targets.\cite{koponen:1997b}

\section{CONCLUSIONS} \label{Conclusions}

In this paper we have considered in detail a two-field continuum description of 
nanopatterning by IBS in the the most general (anisotropic) case of oblique ion incidence. The explicit coupling of the dynamics for diffusing species at the surface with the evolution of the topography assumes exchange between mobile and immobile material at the upper boundary of the latter. This description goes beyond the IBS case and can be employed as a phenomenological formulation of more general phenomena in Surface Science. In the particular case of IBS, this approximation leads to a stationary value for the density of diffusing surface species that is very quickly reached at, as compared with the dynamics of the surface morphology. This fact and the {\em shape} of the effective interface equation are robust to the specifics of the distribution of energy deposition, while these reflect in the values of the model parameters. For the sake of reference, in this work we have been assuming BH behavior for the latter. Under this assumption, we have explored the qualitative 
 properties of the two-field model, and seen that it  indeed provides a more comprehensive framework than previous continuum descriptions fulfilling most of the desiderata formulated in the Introduction. The adoption of a two-field description is not for mere mathematical convenience; rather, it responds to its
enlarged set of physical mechanisms (such as redeposition) entering the constitutive laws in a more natural way ($e.g.$ the coupling between excavation and surface transport, or corrections to BH prediction for the linear velocity of transverse ripple motion). The parameter range for cancellation modes is partially restricted, and important experimentally observed nonlinear behavior, such as wavelength coarsening and non-uniform ripple motion, can be accounted for. Analogous conclusions can be reached at in the cases of normal incidence, or oblique incidence onto rotating targets, that are considered in detail in a forthcoming work.\cite{munoz-garcia:2007b}

Still, some features that remain theoretically unexplained, such as $e.g.$ the lack of pattern formation for IBS of Si at near normal incidence in many experimental settings, may be due to inaccuracies in Sigmund's description of the statistics of energy deposition through collisions inside the target.\cite{davidovitch:2007} Note that, once these are improved upon, the resulting effective parameters could be used in turn as inputs for the local excavation rate \eqref{Ge}. Thus, the two-field model is {\em not} restricted in principle to Sigmund's Gaussian statistics. Moreover, through the mentioned natural coupling between transport and topography, and through the incorporation of redeposition effects, it already provides (albeit, admittedly, in a simplified form that is susceptible to refinement) description of additional material rearrangement due to ion impingement, that currently seems necessary for a theoretical description of IBS with an improved predictive power.\cite{
 kalyana:2008}


\begin{acknowledgments}
We thank R.\ Gago and L.\ V\'azquez for discussions. This work has been
partially supported by MEC (Spain), through Grants No.\ FIS2006-12253-C06-01 and No.\ FIS2006-12253-C06-06, by CAM (Spain) through Grant No.\ S-0505/ESP-0158, and by JCCM through Grant No.\ PAC-05-005.
\end{acknowledgments}

\appendix

\section{Sigmund's energy distribution} \label{app.A}

In [\onlinecite{makeev:2002}], the most general equation of motion for the
local surface velocity reads
\begin{multline}\label{TNLeqh}
\partial_t h=-v_0 + \gamma_x \partial_x h +\Omega_1 \partial^3_x h+\Omega_2 \partial_x\partial_y^2 h \\
+ \left[\xi_x \left(\partial_x h\right) +\nu_x\right]\left(\partial^2_x
h\right)+\left[\xi_y \left(\partial_x h \right)+\nu_y\right] \left(\partial^2_y
h\right) \\
-D_{xx}\partial^4_x h-D_{yy}\partial^4_y h-D_{xy}\partial^2_x\partial_y^2
h+\frac{\lambda_x}{2}\left(\partial_x
h\right)^2+\frac{\lambda_y}{2}\left(\partial_y h\right)^2 ,
\end{multline}
where all coefficients are functions \cite{makeev:2002} of physical parameters
such as $\theta$, $E$, $\Phi$, and the characteristics of Sigmund's Gaussian
energy distribution, such as the average penetration depth, $a$, and the
lateral widths of the distribution $\sigma$ and $\mu$. In particular, the
coefficients $D_{ij}$ correspond to the so-called ion-induced effective
smoothing terms whose terms have the same shape as those characteristic of
surface diffusion, but are of a mere ``geometric'' origin related with
describing the surface height at sufficiently high order in a Taylor expansion
in height derivatives. For this (standard) choice, the values of coefficients
$\alpha_{ikl}$ in the excavation rate $\Ge$ in (\ref{Ge}) are
\begin{align}\label{relacion_Ge}
\alpha_0&=v_0,\quad \alpha_{1x}=-\gamma_x/v_0,\quad
\alpha_{2x,y}=-\nu_{x,y}/v_0,\nonumber \\
\alpha_{3x}&=-\Omega_1/v_0,\quad \alpha_{3y}=-\Omega_2/v_0,\quad \alpha_{4ij}=
-D_{ij}/v_0, \nonumber \\
\alpha_{5x,y}&=-\xi_{x,y}/v_0,\quad \alpha_{6x,y}=-\lambda_{x,y}/v_0 .
\end{align}

\section{Formulae for Sec.\ \ref{linear}} \label{app.AA}

\begin{align}
    &a=\gamma_0+ \sum_{j=x,y} \left[ D+R_{eq}\gamma_0 \left(\gamma_{2j}-\epsilon \alpha_{2j} + \epsilon  \sum_{i=x,y} \alpha_{4ij}k_i^2 \right)  \right] k^2_j , \label{ap_a} \\
    &b=\epsilon \gamma_0 R_{eq} k_x \left(\alpha_{1x} - \sum_{j=x,y} \alpha_{3j} k_j^2\right) , \label{ap_b} \\
    &c=\gamma_0 R_{eq} D (k_x^2+k_y^2) \sum_{j=x,y} \left(\gamma_{2j}-  \epsilon \alpha_{2j} +\epsilon  \sum_{i=x,y} \alpha_{4ij}k_i^2 \right)
    k_j^2 \nonumber\\
     &+\epsilon \phi \gamma_0^2 R_{eq}  \sum_{j=x,y} \left(-  \alpha_{2j} +\sum_{i=x,y} \alpha_{4ij}k_i^2 \right) k_j^2 , \label{ap_c}\\
    &d=\epsilon \gamma_0 R_{eq} k_x \left[D (k_x^2+k_y^2)+\phi \gamma_0\right] \left(\alpha_{1x} - \sum_{j=x,y} \alpha_{3j} k_j^2\right) . \label{ap_d}
\end{align}

\section{Ripple wavelength and orientation} \label{app.B}

In this appendix, we determine the ripple orientation and wavelength
within linear theory. Our starting point is Eq.\ \eqref{ReW} which, neglecting
$O(\epsilon^2)$ terms, can be rewritten as
\begin{equation}  \label{ReW2}
 {\rm Re} (\omega_{\mathbf{k}})= -\nu_x k_x^2 - {} \nu_y k_y^2 - \mathcal{K}_{xx}
 k_x^4 - \mathcal{K}_{yy} k_y^4 - \mathcal{K}_{xy} k_x^2 k_y^2,
\end{equation}
where
\begin{align} \label{nus}
\nu_x&=-\epsilon \phi \gamma_0 R_{eq} \alpha_{2x} ,\quad
\nu_y=-\epsilon \phi \gamma_0 R_{eq} \alpha_{2y} \\
\mathcal{K}_{xx}&=D R_{eq} \gamma_{2x} \label{Dxx} \\
 &+ \epsilon R_{eq} \gamma_0 \left[\phi \alpha_{4xx} - \Big(\frac{\bar{\phi}D}{\gamma_0}-\phi
R_{eq}\gamma_{2x}\Big) \alpha_{2x}\right], \nonumber \\
\mathcal{K}_{yy}&=D R_{eq} \gamma_{2y} \label{Dyy} \\
 &+ \epsilon R_{eq} \gamma_0 \left[\phi \alpha_{4yy} - \Big(\frac{\bar{\phi}D}{\gamma_0}-\phi
R_{eq}\gamma_{2y}\Big) \alpha_{2y}\right], \nonumber \\
\mathcal{K}_{xy}&= D R_{eq} (\gamma_{2x}+\gamma_{2y}) +\epsilon R_{eq} \gamma_0
\Big[2\phi \alpha_{4xy} \label{Dxy} \\
 &- \left. \frac{\bar{\phi}D}{\gamma_0}(\alpha_{2x}+\alpha_{2y})
 +\phi R_{eq}(\gamma_{2x}\alpha_{2y}+\gamma_{2y}\alpha_{2x}) \right]. \nonumber
\end{align}
Note that the above $O(k^4)$ form is a consequence of our long wavelength
approximation to $\omega_{\mathbf{k}}$. However, the precise shape of the
coefficients is sensitive to the order considered in the (independent)
expansion in powers of $\epsilon$, resulting in Eqs.\ \eqref{nus}-\eqref{Dxy}.
Thus, for instance, given that $\gamma_{2x}$ and $\gamma_{2y}$ are positive,
$\mathcal{K}_{xx}$, $\mathcal{K}_{yy}$, $\mathcal{K}_{xy}$ are also always
positive to $O(\epsilon^0)$. However, the signs of their $O(\epsilon)$
contributions can change with experimental conditions, see $e.g.$
[\onlinecite{makeev:2002}].

The experimentally observed pattern is oriented along the direction which
yields the maximum value of the real part of the dispersion relation, and its
wavelength is associated to the wave vector, $\mathbf{k^\ell}=(k_x^\ell, k_y^\ell)$,
which maximizes Eq.\ \eqref{ReW2}. This vector must verify
\begin{equation}\label{min}
\frac {\partial \Rwm}{\partial k_x}\left(\mathbf{k}^\ell \right)= \frac {\partial \Rwm}{\partial k_y}\left(\mathbf{k}^\ell \right)=0. 
\end{equation}
These conditions have the following independent solutions
\begin{align}\label{kas}
    \mathbf{k_0}&=\left(0,0\right),\,
    \mathbf{k_1}=\left(\sqrt \frac{{} -\nu_x}{2\mathcal{K}_{xx}} ,0\right),\,
    \mathbf{k_2}=\left(0, \sqrt
\frac{{} -\nu_y}{2\mathcal{K}_{yy}}\right),\nonumber \\
\mathbf{k_3}&=\left(\sqrt \frac{2 \nu_x\mathcal{K}_{yy}- {} \nu_y
\mathcal{K}_{xy}}{\mathcal{K}_{xy}^2 - 4 \mathcal{K}_{xx}\mathcal{K}_{yy}} ,
\sqrt \frac{2 {} \nu_y \mathcal{K}_{xx}- {} \nu_x
\mathcal{K}_{xy}}{\mathcal{K}_{xy}^2 - 4 \mathcal{K}_{xx}\mathcal{K}_{yy}}
\right).
\end{align}
The solution $\mathbf{k}_3$ exists provided the arguments of the square roots
of its two components are positive, and $\mathcal{K}_{xy}^2 - 4
\mathcal{K}_{xx}\mathcal{K}_{yy}\neq 0$. Otherwise, the only solutions to
\eqref{min} are $\mathbf{k_0}$, $\mathbf{k_1}$, and $\mathbf{k_2}$. Moreover,
as noted above, for large angles of incidence, $\nu_x$ is positive and
$\mathbf{k_1}$ is not defined.

Note, wave vector $\mathbf{k}_3$ implies a surface morphology with a
periodicity that is aligned {\em neither} with the $x$ {\em nor} with the $y$
directions (``oblique ripples''). In order to study this solution, we write
\begin{align}
    {\rm Re} (\omega_{\mathbf{k_1}}) - {\rm Re}
(\omega_{\mathbf{k_3}}) &= \frac{\left(\mathcal{K}_{xy}{}\nu_x
-2\mathcal{K}_{xx}{}
\nu_y\right)^2}{4\mathcal{K}_{xx}\left(\mathcal{K}_{xy}^2-4\mathcal{K}_{xx}
\mathcal{K}_{yy}
\right)}, \label{k13} \\
 {\rm Re} (\omega_{\mathbf{k_2}}) - {\rm Re}
(\omega_{\mathbf{k_3}}) &= \frac{\left(\mathcal{K}_{xy}{}
\nu_y-2\mathcal{K}_{yy}{}
\nu_x\right)^2}{4\mathcal{K}_{yy}\left(\mathcal{K}_{xy}^2-4\mathcal{K}_{xx}
\mathcal{K}_{yy} \right)},\label{k23}
\end{align}
so that the signs of \eqref{k13}, \eqref{k23} are given by that of
\begin{equation}
\mathcal{K}_{xy}^2- 4 \mathcal{K}_{xx}\mathcal{K}_{yy} \label{K} .
\end{equation}
Moreover, straightforward algebra shows that the sign of the determinant of the
Hessian matrix evaluated at $\mathbf{k}=\mathbf{k}_3$ is opposite to that of
\eqref{K}. In summary,
\begin{enumerate}
\item[$(i)$] For positive values of \eqref{K}, $\mathbf{k}_3$ is a saddle
point, so that the absolute maximum of \eqref{ReW2} takes place either at
$\mathbf{k}_1$ or at $\mathbf{k}_2$.

\item[$(ii)$] For negative values of \eqref{K}, $\mathbf{k}_3$ provides the
absolute maximum of \eqref{ReW2}.
\end{enumerate}
These results are valid for any value of $\epsilon$. To order $O(\epsilon^0)$,
it is easy to see from \eqref{Dxx}-\eqref{Dxy} that
$\mathcal{K}_{xy}=\mathcal{K}_{xx}+\mathcal{K}_{yy}$, so that \eqref{K} equals
$(\mathcal{K}_{xx}-\mathcal{K}_{yy})^2>0$ and condition $(i)$ above holds. This
is the situation that occurs in most of the physical systems we will be
considering, due to the smallness of the corresponding values of $\epsilon$.
Higher order corrections are sensitive to high order details of the
distribution of energy deposition. Thus, the sign of the $O(\epsilon)$ term in
\eqref{K} is given, for isotropic thermal surface diffusion
($\gamma_{2x}=\gamma_{2y}$), by the sign of
$2\alpha_{4xy}-\alpha_{4xx}-\alpha_{4yy}$. Hence, for specific choices of these
effective smoothing coefficients it is conceivable that oblique ripples occur
in our two field model for large $\epsilon$ values, but we will not consider
such situations in what follows.

In order to decide which of the remaining solutions provide the absolute
maximum of ${\rm Re} (\omega_{\mathbf{k}})$, we finally substitute the wave
vectors given by \eqref{kas} into Eq.\ \eqref{ReW2}; we obtain simply
\begin{equation}
{\rm Re} (\omega_{\mathbf{k_0}})=0,\quad {\rm
Re}(\omega_{\mathbf{k_1}})=\frac{{} \nu_x^2}{4\mathcal{K}_{xx}},\quad
{\rm Re}(\omega_{\mathbf{k_2}})=\frac{{} \nu_y^2}{4\mathcal{K}_{yy}}.\nonumber 
\end{equation}
Further discussion on the final orientation of the ripple structure can be
found in the main text in Sec.\ \ref{linear_inst}.

\section{Multiple-scale analysis} \label{app.C}

In this appendix, we provide the details for the derivation of the effective
interface equation \eqref{eq.ero}. The setting is provided by formulae
\eqref{eq.Rb} to \eqref{Geepsilon}. For further convenience, we obtain a useful
expression through addition of \eqref{eq.Rb} to \eqref{eq.hb}
\begin{equation}\label{eq.Rh}
  \epsilon^{3/2} \partial_{T_1} \widetilde{h} + \epsilon^{2} \partial_{T_2} \widetilde{h}
= -\phi \widetilde{{\Gamma}}_{ex} + \epsilon D \nabla^2 \widetilde{R} -
\epsilon^{3/2} \partial_{T_1} \widetilde{R} - \epsilon^{2} \partial_{T_2}
\widetilde{R}.
\end{equation}

We will introduce the expansions \eqref{expansionR} and \eqref{expansionh} into
 \eqref{eq.Rnew} and \eqref{eq.Rh} and solve order by order in powers of $\epsilon^{1/2}$.

\paragraph*{Order $\epsilon^0$:}
To this order, as provided by Eq.\ \eqref{eq.Rnew}, there is no contribution and we obtain
\begin{equation}\label{R0}
R_0=0.
\end{equation}
This means that the most important contribution to $\widetilde{R}$ vanishes near the instability threshold. Hence, as we already noted when we obtained the planar solution, $R$ will be only slightly altered from its planar value.

\paragraph*{Order $\epsilon^{1/2}$:}
Again, there are no contributions to this order and from \eqref{eq.Rnew} we
obtain
\begin{equation}\label{R1}
R_1=0.
\end{equation}

\paragraph*{Order $\epsilon^1$:}
At this order, Eq.\ \eqref{eq.Rnew} reads
\begin{equation}\label{R2}
  R_2=-R_{eq}\nabla \cdot (\underline{\gamma_2} \nabla
  h_0),
\end{equation}
which yields the first correction to the expansion of $\widetilde{R}$ and
depends on the curvatures of $h_0$. As anticipated in the main text, $R_n$
contributions indeed depend of lower order $h_m$ terms.


\paragraph*{Order $\epsilon^{3/2}$:}
From \eqref{eq.Rnew} to this order we obtain
\begin{equation}\label{R3}
  R_3= - R_{eq} \nabla \cdot
  (\underline{\gamma_2} \nabla h_1)+ \bar{\phi} R_{eq} \alpha_{1x} \partial_X h_0.
\end{equation}
We can substitute the previous values for the expansion of $\widetilde{R}$ into
\eqref{eq.Rh} to finally obtain
\begin{equation} \label{Dth0}
  \partial_{T_1}h_0= - \phi \gamma_0 R_{eq} \alpha_{1x} \partial_X h_0,
\end{equation}
which allows us to formally solve for $h_0$. Note that this equation provides
in-plane propagation as the leading contribution to $\widetilde{h}$ in the slow
time scales, with the same velocity as predicted by the imaginary part of the
linear dispersion relation, Eq.\ \eqref{ImW}.

\paragraph*{Order $\epsilon^{2}$.}
Following the previous scheme, from \eqref{eq.Rnew} we have
\begin{equation}\label{R4}
R_4=- R_{eq} \nabla \cdot (\underline{\gamma_2} \nabla h_2)+\frac{ \bar{\phi}}
{ \gamma_0} \widetilde{\Gamma}_{ex}(\epsilon^2)+ \frac{D}{\gamma_0} \nabla^2
R_2,
\end{equation}
where we have used $\widetilde{\Gamma}_{ex}(\epsilon^2)$ to denote the order
$\epsilon^2$ contribution of $\widetilde{\Gamma}_{ex}$, given by
\begin{equation}
  \widetilde{\Gamma}_{ex}(\epsilon^2)= \gamma_0 R_{eq} \left[\alpha_{1x} \partial_X h_1+
\nabla \cdot (\underline{\alpha_2} \nabla h_0) + \nabla h_0 \cdot
(\underline{\alpha_6} \nabla h_0) \right].
\end{equation}
From \eqref{eq.Rh} we obtain again a closed evolution equation for
$\widetilde{h}$ to order $\epsilon^2$, namely,
\begin{equation} \label{Dth1}
  \partial_{T_1}h_1+\partial_{T_2}h_0= -\phi \widetilde{\Gamma}_{ex}(\epsilon^2) +D
\nabla^2 R_2.
\end{equation}
As we can see from the previous results, to obtain the temporal derivatives of
$\widetilde{h}$ to order $\epsilon^{n/2}$ we need to know $\widetilde{R}$ to
order  $R_{n-2}$. Since we already know the value of the expansion of
$\widetilde{R}$ up to $R_4$, we can obtain a closed evolution equation for
$\widetilde{h}$ to order $\epsilon^{3}$.

\paragraph*{Order $\epsilon^{5/2}$:}
From \eqref{eq.Rh} we have
\begin{equation} \label{Dth2}
  \partial_{T_1}h_2+\partial_{T_2}h_1= -\phi \widetilde{\Gamma}_{ex}(\epsilon^{5/2})
+D \nabla^2 R_3 - \partial_{T_1}R_2,
\end{equation}
where we have again used $\widetilde{\Gamma}_{ex}(\epsilon^{5/2})$ to denote
the $\epsilon^{5/2}$ order contribution to $\widetilde{\Gamma}_{ex}$. The time
derivative of $R_2$ can be computed by making use of Eqs.\ \eqref{R2} and
\eqref{Dth0}, to get
\begin{equation}
  \partial_{T_1} R_2= \phi \gamma_0 R_{eq}^2 \alpha_{1x} \nabla \cdot
  (\underline{\gamma_2} \partial_X h_0).
\end{equation}

\paragraph*{Order $\epsilon^{3}$:}
Similarly to the previous step, from \eqref{eq.Rh} we now have
\begin{equation}
    \partial_{T_1}h_3+\partial_{T_2}h_2= -\phi
\widetilde{\Gamma}_{ex}(\epsilon^{3}) +D \nabla^2 R_4 - \partial_{T_1}R_3 -
    \partial_{T_2}R_2. \label{Dth3}
\end{equation}
Noting that $\widetilde{\Gamma}_{ex}(\epsilon^{n})$ do {\em not} depend
explicitly on $\widetilde{R}$, and since we (formally) know the values of
$R_2$, $R_3$, and $R_4$ as functions of terms of the expansion of
$\widetilde{h}$, we can finally obtain a closed equation for the evolution of
$\widetilde{h}$ up to order $\epsilon^3$. To this end, using the relation
between the slow temporal variables given by \eqref{Dt}, we can write the time
derivative of the expansion of $\widetilde{h}$, Eq.\ \eqref{expansionh}, as
\begin{multline}
  \partial_t \widetilde{h}  = \epsilon^{3/2} \partial_{T_1} h_0 + \epsilon^2
  \left(\partial_{T_2} h_0 + \partial_{T_1} h_1 \right)  \\
   + \epsilon^{5/2} \left( \partial_{T_2} h_1 + \partial_{T_1} h_2 \right) +
\epsilon^3 \left( \partial_{T_2} h_2 +
  \partial_{T_1} h_3 \right). \label{desarrolloh}
\end{multline}
Since all terms in \eqref{desarrolloh} are known, substituting Eqs.\
\eqref{Dth0}, \eqref{Dth1}, \eqref{Dth2}, and \eqref{Dth3} into
\eqref{desarrolloh}, and using \eqref{Dth1} in order to simplify the ${\cal
O}(\epsilon^3)$ contribution in \eqref{desarrolloh}, we obtain
\begin{widetext}
\begin{multline}\label{hXY}
\partial_t \widetilde{h} = \epsilon^{3/2} \left\{ - \phi \gamma_0 R_{eq} \alpha_{1_x}
\partial_X \widetilde{h} \right\} + \epsilon^2 \left \{ -\phi\gamma_0 R_{eq}
\left[\nabla \cdot (\underline{\alpha_2} \nabla \widetilde{h})+ \nabla
\widetilde{h} \cdot (\underline{\alpha_6} \nabla \widetilde{h}) \right] - D
R_{eq}\nabla^2 [\nabla \cdot
(\underline{\gamma_2} \nabla \widetilde{h})] \right\}\\
+\epsilon^{5/2} \left\{- \phi \gamma_0 R_{eq} \left[\partial_X \nabla \cdot
(\underline{\alpha_3} \nabla \widetilde{h})+ (\partial_X \widetilde{h}) \nabla
\cdot (\underline{\alpha_5} \nabla \widetilde{h})\right] + \alpha_{1x} \nabla
\cdot \left[\left(\bar{\phi} D R_{eq} \mathbb{I}-\phi \gamma_0 R_{eq}^2
\underline{\gamma_2}\right) \nabla \partial_{X} \widetilde{h} \right] \right\}\\
+ \epsilon^3 \left\{- \phi \gamma_0 R_{eq} \nabla \cdot (\underline{\alpha_4}
\nabla \nabla^2 \widetilde{h}) + \nabla \cdot \left[\left( \bar{\phi} D R_{eq}
\mathbb{I}-\phi \gamma_0 R_{eq}^2 \right) \underline{\gamma_2}\right] \nabla
\left[ \nabla \cdot (\underline{\alpha_2} \nabla \widetilde{h})+ \nabla
\widetilde{h} \cdot (\underline{\alpha_6} \nabla \widetilde{h}) \right] +
\bar{\phi} \phi \gamma_0 R_{eq}^2 \alpha_{1x}^2 \partial_X^2 \widetilde{h}
\right\},
\end{multline}
\end{widetext}
where $\mathbb{I}$ stands for the $2\times 2$ identity matrix, we have again
employed the expansion of $\widetilde{h}$, Eq.\ \eqref{expansionh}, and we have
neglected sixth order derivatives of $\widetilde{h}$.

In order to compare with the original model and the dispersion relation
obtained in Sec.\ \ref{linear}, we return to the original variables. Recalling
that $h=h^p+\widetilde{h}$, $X=\epsilon^{1/2}x$, $Y=\epsilon^{1/2}y$, and
$\epsilon=\alpha_0/(\gamma_0 R_{eq})$, we finally obtain equations
\eqref{eq.ero} and \eqref{relacion_eq.ero} of the main text.

\bibliography{bibliografia}

\end{document}